\documentclass[useAMS,usenatbib,usegraphicx]{mn2e}

\usepackage{amssymb,amsmath}

\newcommand{\kms}{\hbox{km s$^{-1}$}}
\newcommand{\re}{\hbox{${\rm R}_{\rm e}$}}

\newcommand{\msun}{\hbox{M$_{\odot}$}}
\newcommand{\reduceme}{\mbox{R\raisebox{-0.35ex}{E}D%
\hspace{-0.05em}\raisebox{0.85ex}{uc}\hspace{-0.90em}%
\raisebox{-.35ex}{{m}}\hspace{0.05em}E}}

\title[Radial IMF variations in ETGs]{Radial variations in the stellar initial mass function of early-type galaxies}
\author[Mart\'in-Navarro et al.]{Ignacio Mart\'in-Navarro$^{1,2}$\thanks{E-mail: imartin@iac.es}, Francesco La Barbera$^{3}$, Alexandre Vazdekis$^{1,2}$, 
\newauthor
Jes\'us Falc\'on-Barroso$^{1,2}$ and  Ignacio Ferreras$^{4}$\\ 
$^{1}$Instituto de Astrof\'isica de Canarias, E-38200 La Laguna, Tenerife, Spain\\
$^{2}$Departamento de Astrof\'isica, Universidad de La Laguna, E-38205 La Laguna, Tenerife, Spain\\
$^{3}$INAF - Osservatorio Astronomico di Capodimonte, Napoli, Italy\\
$^{4}$Mullard Space Science Laboratory, University College London, Holmbury St Mary, Dorking, Surrey RH5 6NT, UK}

\begin{document}

\date{ }

\pagerange{\pageref{firstpage}--\pageref{lastpage}} \pubyear{2014}

\maketitle
\label{firstpage}
\begin{abstract}
The hypothesis of a universal initial mass function (IMF) -- motivated
by  observations  in  nearby  stellar  systems --  has  been  recently
challenged by  the discovery of  a systematic variation  of the IMF with the
central velocity dispersion,  $\sigma$, of early-type galaxies (ETGs),
towards an  excess of low-mass  stars in high-$\sigma$  galaxies. 
This trend has been derived so far from integrated spectra, and
remains unexplained at present.  To test whether such trend depends on
the {\sl local} properties within a galaxy, we have obtained new,
extremely deep, spectroscopic data, for three nearby ETGs, two
galaxies with high $\sigma$ ($\sim 300$~$\rm km \, s^{-1}$), and one
lower mass system, with $\sigma \sim 100 \, \rm km \, s^{-1}$.  From the
analysis of IMF-sensitive spectral features, we find that the IMF
depends significantly on galactocentric distance in the massive ETGs,
with the enhanced fraction of low-mass stars f mostly confined  to their central
regions. In contrast, the low-$\sigma$ galaxy does not show any
significant radial gradient in the IMF, well described by a shallower
distribution, relative to the innermost regions
of massive galaxies, at all radii. Such a result indicates that
the IMF should be regarded as a local (rather than global) 
property, and suggests a significant difference between the formation
process of the core and the outer regions of massive ETGs.
\end{abstract}

\begin{keywords}
galaxies: formation -- galaxies: evolution -- galaxies: elliptical --
galaxies: fundamental parameters
\end{keywords}

\section{Introduction} 

The stellar Initial Mass Function (IMF) characterises the distribution
of stellar masses at birth in star forming regions. The IMF is
therefore a crucial ingredient of galaxy formation and evolution.  It
sets the mass-scale of galaxies, determining their (stellar)
mass-to-light ratio, and drives stellar feedback as well as chemical
enrichment into the ISM.  While resolved stellar population studies
support the idea of an invariant IMF in environments with quite
different local properties such as metallicity or density
\citep{kroupa,bastian,kroupa13}; recent studies of early-type galaxies
(ETGs), based on both dynamics
\citep[][]{cappellari,thomas11,wegner12,dutton13,tortora13} and
stellar populations
\citep[][]{saglia,cenarro,vandokkum,spiniello12,ferreras,labarbera}
have found that the IMF varies with galaxy mass.  {  The same result
  has also been obtained by a combination of gravitational lensing and
  dynamical studies \citep[][but see \citealt{smith13}]{treu,auger}.}  In 
particular, dynamical studies have found a
significant increase of the stellar mass-to-light ratio, with respect
to that expected for a ``standard'', Milky-Way-like IMF, towards
high-mass systems. The analysis of gravity-sensitive features in the
integrated spectra of ETGs has revealed that this trend in the
``normalization'' of the IMF is driven by an increase of the
fraction of dwarf-to-giant stars, i.e. a change towards steeper IMF slopes
with higher velocity dispersion \citep[][hereafter
  LB13]{cenarro,jesus,cappellari,ferreras,labarbera}.  {  Although
  some discrepancies can be found between dynamical and stellar
  population studies, the agreement is remarkable, considering the
  fundamental differences of both approaches \citep{smith}.}

One should notice that the results from stellar population studies are
derived from integrated spectra, therefore correspond mostly to the
bright central regions of ETGs. Therefore, the question of whether
radial variations {\sl within} a galaxy occur, follows naturally. Are
these variations in the IMF driven by a large-scale (e.g.  galaxy
mass) or a local property (e.g. local velocity dispersion) ?  Despite
the importance of this question to constrain the overall picture of
galaxy formation and evolution, no reliable spatially-resolved
measurement of the IMF has been performed so far.  Only a few early
attempts tried to investigate this issue. \citet{carter} found strong
radial gradients of Na and TiO spectral features in ETGs, interpreting
it as the contribution from metal-rich populations of dwarf stars,
concentrated towards the centres of the most massive galaxies.  This
interpretation was dismissed by \citet{cohen,hardy88} and
\citet{delisle92}, who suggested instead metallicity alone as the
driver of these radial trends.  Due to the lack of accurate stellar
population models and high-quality data at the time, these pioneering
attempts remained inconclusive.

Over twenty years later, in the present work, we show that after
considerable improvement in the state-of-the-art stellar population
synthesis models, analysis tools, observational facilities, and
instrumentation, we are now able to address the issue of a radial
variation of the IMF in unresolved stellar populations.  Targeting a
set of optical and Near-Infrared (NIR) gravity-sensitive spectral
features in two high-$\sigma$ ($\rm \sim 300 \, km \, s^{-1}$) and one
low$-\sigma$ ($\rm \sim 100 \, km \, s^{-1}$) ETGs, we find that
variations in the IMF of these systems should be regarded as a local
property.

The outline of the paper is the following.  In Section~2, we describe
the sample of ETGs and data reduction.  Section 3 presents the stellar
population analysis. Section~4 shows the main results of the present
work, i.e.  the IMF trend as a function of galactocentric distance in
ETGs. In Section~5, we present results from a number of tests
performed to assess the robustness of our results.  In Section~6 we
discuss several effects that might mimic an IMF gradient, showing that
none of them can account for all the available data. {  Our results are
  discussed in Section~7, and conclusions are given in Section~8.  Further
  material is presented in Appendix~B, to address possible technical
  issues of the stellar population analysis.}

\section{Sample and data reduction}
\label{sec:data}
{  We obtained deep spectroscopic data of three nearby early-type
  galaxies, with different masses and different radial profiles of
  velocity dispersion. NGC\,4552 and NGC\,5557 are massive ETGs, with
  a velocity dispersion $\sigma\sim 300$\,\kms \ in their
  centre~\citep{kine}, whereas NGC\,4387 is a lower-mass system with a
  central value of $\sigma \sim 100$\,\kms .  The effective radii are, 
  18.\arcsec85, 36.\arcsec31 and 28.\arcsec84 for NGC\,4387, NGC\,4552 and 
  NGC\,5557, respectively \citep{atlas3d}. Because the spectra of 
  NGC~5557 turned out to be significantly
  affected by telluric absorption, we analyzed its data following a
  different approach than for NGC\,4552 and NGC\,4387.  While the
  kinematics of NGC\,5557 is presented in Sec.~\ref{sec:kine},
  together with that for the other two galaxies, the analysis of the IMF is
  presented separately, in Appendix~\ref{sec:NGC5557}. Throughout the paper, we
  refer to NGC\,4387 and NGC\,4552 as our reference low- and high-mass
  ETGs. We notice that both NGC\,4387 and NGC\,4552 (as well as
  NGC\,5557) are not peculiar in terms of either kinematics or stellar
  population properties \citep{atlas3d}.}

The observations were carried out at the 10.4m Gran Telescopio
Canarias, with the OSIRIS \citep{osiris} spectrograph.  The spectra
cover the range from 4500 to 10000\,\AA\, with a resolution of
$R=$~2500 for $\lambda<$~6000\,\AA, and $R=$ 1,000 otherwise. The slit
used has 1.\arcsec0 $\times$ 7.\arcmin8 along the spectral and spatial
directions, respectively. The wide wavelength range allows us to
target IMF-sensitive absorption features that are sensitive to
different chemical species (mostly, Na, Ca, and Ti). This is crucial
to distinguish between the effect of a varying IMF and other effects
possibly affecting gravity-sensitive indices, such as variations of
elemental abundance ratios. Each target was observed for 1.5 hours,
allowing us to achieve the high signal-to-noise ($S/N$) ratio required
to analyze the IMF effect on absorption features.

Data reduction was done with the {\reduceme} package~\citep{reduceme},
allowing for a careful propagation of different sources of uncertainties
during the reduction. This is crucial for our purposes, as we aim to
explore different galactocentric regions -- up to about one effective
radius -- where systematic effects might be potentially important
because of the low surface brightness level. We performed the usual
spectroscopic reduction steps, including bias subtraction,
flat-fielding, cosmic ray cleaning, sky subtraction, C-distortion
correction, wavelength calibration (better than $\Delta\lambda \sim 8$
\%), S-distortion correction, extinction correction and flux
calibration.

In order to measure the radial variation of the IMF, for each galaxy
we produced averaged spectra in different radial bins.  To this
effect, we first corrected, line-by-line, the two-dimensional reduced
spectrum to the rest-frame (removing both systemic and rotation
velocity), and convolved each spectrum (i.e. each line) to match the
maximum velocity dispersion of the galaxy (see \S\ref{sec:kine} for
details on the determination of the kinematics).  Then, we defined
radial bins adaptively, summing up the spectra along the slit position
incrementally, until a target $S/N$ \break ratio $\geq$ 100 per \AA
\ was reached for each bin.

\section{Analysis}

\subsection{Kinematics}\label{sec:kine}

\begin{figure}
\begin{center}
\includegraphics[width=8.5cm]{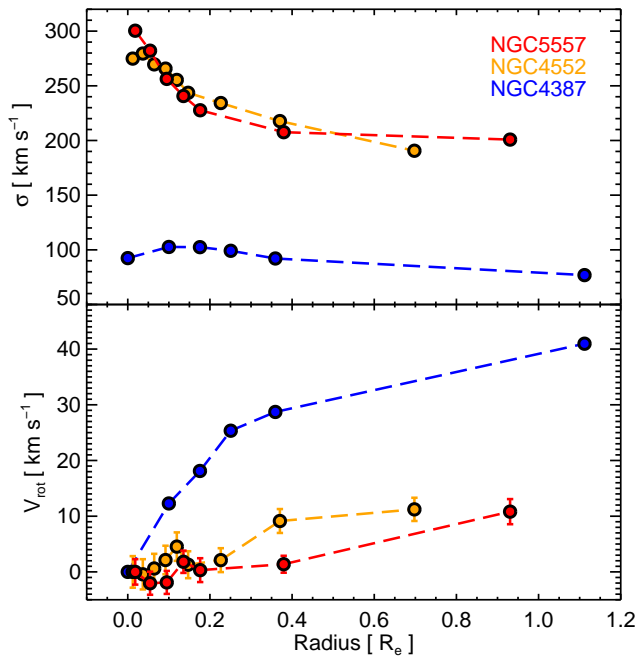}
\end{center}
\caption{{  Radial gradients of velocity dispersion (top) and rotation velocity (bottom), from our 
GTC-OSIRIS observations, i.e.
for NGC~4552, NGC~4387, and the additional massive galaxy NGC\,5557}. The profiles are plotted against the normalized 
galactocentric distance, R$/$R$_e$. The two massive  galaxies, NGC~5557
  and  NGC~4552,  show a  decreasing  $\sigma$ profile,  while the
  low-mass system, NGC~4387, has an almost  constant velocity
  dispersion profile.  Error  bars in $\sigma$ are  smaller  than  the  symbol
  size. Notice that the radial velocity is more important in the fast-rotator 
  NGC~4387 than in the two massive slow-rotators ({  NGC~4552 and NGC~5557}). 
  The increasing  distance of radial bins
  reflects the signal-to-noise criterion imposed to perform an adaptive 
  binning of the available spectra along the slit spatial direction.  }
\label{fig:sig}
\end{figure}

To measure the radial velocity and velocity dispersion profiles, we
used the software pPXF \citep{ppxf}. The $S/N$ requirement for the
measurement of the kinematics is lower than for the analysis of the
stellar populations. Thus, as explained above, we corrected the
differences in the kinematics of individual spectra within the radial
bins to be used in the study of the stellar population properties. In
Fig.~\ref{fig:sig} we show the profiles of velocity dispersion and
radial velocity as a function of normalized galactocentric distance,
$\rm R/R_e$, {  for all three galaxies observed with GTC-OSIRIS.
  Our measurements are in agreement with previous
  results~\citep{kine}, with the massive galaxy NGC~4552, as well as
  the additional high-mass system NGC\,5557
  (Appendix~\ref{sec:NGC5557}), being slow rotators
  ($\lambda_\mathrm{Re}=$0.049 for both galaxies), and the low-mass
  galaxy NGC~4387 classified as a fast rotator
  ($\lambda_\mathrm{Re}=$0.399)}.

\subsection{Stellar population models}
\label{sec:models}
We analysed the spectra of each galaxy, at each galactocentric
distance, with the \citet{miuscat} extended version of MILES stellar
population models \citep{miles}, hereafter defined as MIUSCAT models.  These
models combine a variety of empirical stellar libraries over the
spectral range $\lambda\lambda 3465-9469$\,\AA, at a nominal
resolution of $2.51$\,\AA\ FWHM, relying on solar-scaled isochrones
with stellar spectra following the abundance pattern of our Galaxy,
i.e.  approximately solar-scaled at solar metallicity.  We consider a
variety of extended MILES Simple Stellar Population (hereafter SSP)
models covering a wide range of ages, from $0.5$ to $14$~Gyr, and
metallicities, from [Z/H]$=-0.4$ to $+0.3$, as well as different IMF
slopes. Notice that we did not use SSPs younger than $0.5$\,Gyr and
with $[Z/H]<-0.4$, as our targeted ETGs are mostly composed of old,
metal-rich, stellar populations.  For metallicities above $+0.22$ --
the maximum MILES value -- SSPs are computed by linear extrapolation
of the available models.  However, our conclusions are not affected at
all by this extrapolation, as only the innermost radial bin of
NGC\,4552 benefits from it (see \S\ref{sec:grid}).  For the present
study, we consider the bimodal IMF \citep{vazdekis96}, consisting of a
power law at high masses, with index $\Gamma_\mathrm{b}$, tapered off
to a constant value at M$<0.6$\,M$_\odot$. As shown in our previous
work (LB13), both bimodal and unimodal -- i.e. single-segment,
Salpeter-like IMFs \citep{Salp:55} -- are able to describe equally
well the gravity sensitive features of ETGs, but bimodal IMFs provide
mass-to-light ratios in better agreement with dynamical
constraints.  For $\Gamma_\mathrm{b} \sim 1.35$, the bimodal IMF
  represents well the canonical, Kroupa-like, IMF
  \citep[e.g.][]{mw}. Lower (higher) values of
$\Gamma_\mathrm{b}$ correspond to top- (bottom-)heavy distributions,
with an excess of high- (low-) mass stars.  The lower and upper mass
cutoffs of the IMFs are set to $0.1$ and $100$\,M$_\odot$,
respectively, while the IMF slope, $\Gamma_\mathrm{b}$, is allowed to
vary from $\Gamma_\mathrm{b}=0.3$ to $3.3$, where MILES models provide
robust predictions \citep{miuscat}.

\subsection{Fitting process} \label{sec:fit}
{  
For each spectrum, i.e. each galaxy and radial bin, we infer the best-fitting slope  of 
the bimodal IMF, $\Gamma_b$, by minimising the expression:
\begin{align}
\chi^2(\Gamma_\mathrm{b}, {\rm [Z/H]}, \delta {\rm Ti } )= \left[
  \frac{Age(\Gamma_\mathrm{b})-Age_M}{\sigma_{Age}} \right]^2+ \nonumber \\
  \sum_i \left[
  \frac{(EW_i-C_{\alpha,i} \cdot [\alpha/{\rm Fe}] - \Delta_{\rm Ti,i} \cdot \delta {\rm Ti})-EW_{M,i} }{\sigma_{EW_i}} \right]^2,
\label{eq:method}
  \end{align} 
where $Age(\Gamma_\mathrm{b})$ is the ($\Gamma_\mathrm{b}$-dependent)
estimate of the age, obtained from spectral fitting (see
Sec.~\ref{sec:age} for details); $EW_i$ are the measured equivalent
widths (i.e.  line-strengths) for a selected set of spectral features
(see Sec.~\ref{sec:indices}); and $Age_M$ and $EW_{M,i}$ are the Age
and equivalent widths of (MILES) SSP models; $C_{\alpha,i}$ is the
observed sensitivity of the $i$-th index to [$\alpha$/Fe] (see
Sec.~\ref{sec:proxi}); $\Delta_{\rm Ti,i}$ is the expected sensitivity
(from \citealt{conroy}, hereafter CvD12, theoretical stellar
population models) to [Ti/Fe] abundance, and $\delta {\rm Ti}$ is a
``residual'' TiO-based abundance correction~\footnote{  Notice,
  also, that $\delta {\rm Ti}$ \textit{is not} a Titanium abundance
  measurement ([Ti/Fe]), since $\delta {\rm Ti}$ strongly depends on
  flux calibration at the level of a few percent (see \S\ref{sec:test}
  and \S\ref{sec:corre}).} \citep[i.e.  not accounted for by the
  $C_{\alpha,i}$ terms, see][]{labarbera}.  The $\chi^2$ minimization
is performed over a range of SSP models, with varying age,
metallicity, and IMF (see Sec.~\ref{sec:models}).  The $\sigma_{Age}$
is a {  rescaled} uncertainty on $Age(\Gamma_\mathrm{b})$, defined
by dividing the 1-sigma error on age ($\sigma'_{Age}$, as estimated
from spectral fitting, see Sec.~\ref{sec:age} below) by the root
square of the number of indices used in the fitting. In practice, this
procedure ensures that the age of the SSP model that minimizes the
$\chi^2$ (Eq.~\ref{eq:method}) is always close to the age estimate
from spectral fitting, i.e. that the age of the preferred SSP model is
{\it not} driven by the IMF-sensitive features (e.g.  TiO1, TiO2, and
Mg4780, whose sensitivity to age, in addition to IMF slope, is also significant;
see Sec.~\ref{sec:indices}).  }

The free fitting parameters are therefore IMF slope
($\Gamma_\mathrm{b}$); total metallicity ([Z/H]); and a Ti-related
correction ($\delta{\rm Ti}$), {  whereas age is a constrained
  fitting parameter (its value being constrained, within
  $\sigma_{Age}$, to the estimate obtained from spectral fitting, for
  each $\Gamma_\mathrm{b}$).  Uncertainties on best-fitting parameters
  are estimated by shifting the $EW_i$'s and $Age(\Gamma_\mathrm{b})$
  (see Eq.~\ref{eq:method}) according to their errors ($\sigma_{EW_i}$
  and $\sigma'_{Age}$, respectively), and repeating the $\chi^2$
  minimization procedure. The uncertainty of a given parameter is 
  bootstrapped from the
  standard deviation of best-fitting repeated estimates.  The term
  $C_{\alpha,i} \cdot$ [$\alpha$/Fe] corrects the observed
  line-strengths to [$\alpha$/Fe]=0, allowing a more direct comparison
  to the reference, solar-scaled, MILES models}. The term $\Delta_{\rm
  Ti,i} \cdot \delta {\rm Ti}$ removes the effect of residual [Ti/Fe]
abundance (and flux calibration effects, see details in
\S\ref{sec:test}) from line-strengths.  The crucial aspects of Eq.~1,
and the rationale beyond all different terms in it, are explained in
the following subsections (\S3.3.$\ast$). We remark that for
NGC\,5557, because of significant contamination from telluric
absorption in the spectra, we have analyzed the spectra {  separately, }
with a different set of indices,  as detailed in
Appendix~\ref{sec:NGC5557}. 


\subsubsection{Spectral indices}
\label{sec:indices}
{  The spectral features we use in the $\chi^2$ minimization
  procedure include the total metallicity indicator
  [MgFe]$^\prime$~\citep{TMB:03}, and the following set of
  gravity-sensitive features: Mg\,4780 \citep{serven}, TiO$_1$
  \citep{trager} ($\lambda$\,$\sim$\,5960\,\AA), TiO$_2$ \citep{trager}
  ($\lambda$\,$\sim$\,6230\,\AA), and Ca$_2$ \citep{cat} ($\lambda$\,$\sim$\,
  8540\,\AA), all of them showing a significant response to a varying
  fraction of low- to high-mass stars in the IMF (LB13).  We note that
  other (gravity-sensitive) NIR Calcium absorption lines, Ca$_1$
  ($\lambda$\,$\sim$\,8500\,\AA ) and Ca$_3$ ($\lambda$\,$\sim$\,8660\,\AA ),
  entering the definition of the Calcium triplet
  (CaT$\equiv$Ca$_1+$Ca$_2+$Ca$_3$; see~\citealt{cat}), are severely
  contaminated by airglow in the spectra of our sample of ETGs, and
  thus are not considered in this paper.  Other well-known
  gravity-sensitive features, including the optical Na doublet NaD
  ($\lambda$\,$\sim$\,5900\,\AA ), and the NIR Na\,I doublet feature
  \citep{Schiavon:00} Na{\sc I}8190 ($\lambda\lambda$\,8183,8195\,\AA),
  are also contaminated by airglow and telluric absorption. However,
  as shown in Appendix~\ref{app:A}, the contamination mostly affects
  the absolute value of the indices, while their radial variation
  (i.e.  the gradient) is robust. Therefore, while we do not include
  Na features in the $\chi^2$ minimization procedure, we compare their
  observed gradients to model gradients, accounting for the radial
  variations of age, metallicity, and IMF as inferred from our
  fiducial set of spectral features (see Appendix~\ref{app:A}).
  Notice that the Na{\sc I}\,8190 feature, together with the Wing-Ford
  band at $\lambda\sim$9900\,\AA, have been actually used by
  \citet{vandokkum} to infer an enhanced population of low-mass stars
  in the centre of elliptical galaxies.  Unfortunately, in the spectra
  of our samples of ETGs, similar issues as for the Na features also
  affect the gravity-sensitive spectral indices, CaH1 ($\lambda \sim
  6380$~\AA ), CaH2 ($\lambda \sim 6850$~\AA ) and aTiO ($\lambda \sim
  5500$~\AA ), recently proposed by \citet{Spiniello2013}. In
  particular, both CaH lines are affected by telluric absorption, CaH1
  is also affected by emission, whereas the aTiO passband encompasses
  a strong sky emission line ($\lambda \sim 5576$\,\AA). {   We
    note that the spectal index, bTiO, of \citet{Spiniello2013} is
    very similar to the \citealt{serven} Mg\,4780 index, that we
    already include in the $\chi^2$ minimization procedure}. In
  Appendix~\ref{app:A}, we show that, in spite of these problems, the
  gradients of CaH2 and aTiO can be safely estimated (the latter with
  large error bars), while for CaH1 the radial gradient is more
  uncertain (being more dependent on the way we perform sky
  subtraction). Nevertheless, we can obtain useful constraints on the
  radial variation of this index.  Therefore, we treat CaH1, CaH2 and
  aTiO in the same way as Na features, i.e.  we do not include them in
  the $\chi^2$ minimization procedure, but we compare instead the
  observed gradients with model predictions for our fiducial set of
  spectral features, breaking the degeneracy between IMF and the other
  stellar population parameters (e.g. elemental abundances).
  
}

\subsubsection{Age determination}
\label{sec:age}
  The age determination of a  stellar population is often performed by
  means  of the  hydrogen  Balmer lines,  whose  line-strengths  might
  replace  the first  term on  the  right hand  side of  Eq.~1 in  the
  $\chi^2$ minimisation  procedure.  In practise,  we do not  use this
  approach here, as (i) H$\beta$ appears significantly contaminated by
  nebular emission  for NGC\,4552, with the  emission correction being
  rather   uncertain   for  this   galaxy   \citep{sarzi};  and   (ii)
  higher-order Balmer lines (i.e.  H$\gamma$ and H$\delta$), which are
  virtually  unaffected by nebular  emission, do  not fall  within the
  observed  spectral range.  

  We constrain instead the age by direct-fitting of each spectrum in
  the wavelength region $\lambda\sim 4600-5600$\,\AA, where prominent
  age and metallicity features are found, using a single SSP
  multiplied by a tenth-degree polynomial. {  We perform the fit
    for each of the ten bimodal IMFs~\footnote{  with slopes
      $\Gamma_\mathrm{b} \! = \{ 0.3, 0.8, 1.0, 1.3, 1.5, 1.8, 2.0,
      2.5, 2.8, 3.0 \} $,} provided by MIUSCAT models
    (Sec.~\ref{sec:models}).  For a given IMF, the best-fitting
    parameters are the age and metallicity of the SSP, as well as the
    polynomial coefficients. Notice that the multiplicative polynomial
    accounts for deviations between the continua of data and models
    (e.g. flux calibration uncertainties in the data).  For each IMF,
    we perform 1000 iterations of the fitting procedure, shifting
    randomly the flux values in the spectra according to their
    uncertainties.  This procedure  provides an age estimate (i.e.
    $Age(\Gamma_\mathrm{b})$ in Eq.~1), along with its
    uncertainty~\footnote{  computed as the standard deviation of
      age estimates from the 1000 iterations},
    $\sigma'_{Age}(\Gamma_\mathrm{b})$, as a function of
    $\Gamma_\mathrm{b}$}. {  Notice that $\sigma_{Age}$ in
    Eq.~\ref{eq:method} is obtained by dividing $\sigma'_{Age}$ by the
    root square of the number of spectral indices fitted, in order to
    give a similar weight to the first term on the right hand side of
    Eq.~1 with respect to the second term (i.e.  the index summation),
    that contains most of the information on the IMF.  
    Although the age can not be derived directly from the
    line-strengths (i.e. Balmer lines, see above), the IMF inference
    is very robust against a variety of age determination methods,
    with different methods providing consistent $\Gamma_b$ gradients
    (see \S\ref{sec:test}).  For NGC\,4387, where the emission
    correction to H$\beta$ is negligible ($<0.03$\,\AA), we verified
    that replacing the $Age$ term in the $\chi^2$ definition (Eq.~1)
    with an additional term in the equivalent widths corresponding to
    the H$\beta$ line, does not change the inferred values of
    $\Gamma_\mathrm{b}$ (also providing very consistent age estimates,
    within $\sim$1\,Gyr\footnote{   The same test was also
      performed on the additional high-mass galaxy, NGC\,5557, for
      which results are only presented in
      Appendix~\ref{sec:NGC5557}. The emission correction for this
      galaxy is negligible, and we find very consistent
      $\Gamma_\mathrm{b}$'s (within the errors, see Appendix) when
      constraining the age by either H$\beta$ or spectral fitting.  },
    to those derived from spectral fitting).  Hence, our procedure is
    robust, and general enough, to constrain the age for all available
    spectra.}

\subsubsection{Chemical abundances and temperature effects} \label{sec:proxi}
  Chemical abundances can  partly mimic the effect of  a varying IMF.
  Moreover, we  want to rely  on empirical (rather  than theoretical),
  solar-scale,  stellar  population  (MILES)  models  to  analyse  the
  spectra. We take  the abundance issue into account  in two ways.

  First, using the semi-empirical correction  procedure described in LB13, we
  correct the observed line  strengths to solar scale. The corrections
  to line  strengths are the terms  $C_{\alpha,i} \cdot $[$\alpha$/Fe]
  in  Eq.~1,   where  $C_{\alpha,i}$  are   semi-empirical  correction
  coefficients,  given  by the  slopes  of  the  correlations of  line
  strengths with  [$\alpha$/Fe], at fixed velocity  dispersion -- from
  stacked spectra  of ETGs in  the Sloan Digital Sky  Survey (see~LB13
  for  details).   For each  spectrum,  we
  estimate   [$\alpha$/Fe]    (and   plug   it    into   $C_{\alpha,i}
  \cdot$[$\alpha$/Fe])  by a  solar-scale  proxy, $[Z_{\rm  Mg}/Z_{\rm
      Fe}]$, defined as the  difference of the metallicities estimated
  with  either Mg  or Fe  lines, for  a fixed  Kroupa IMF,  with MILES
  models.  Notice that  the [$\alpha$/Fe] correction procedure assumes
  that the $C_{\alpha,i}$'s, derived in galaxy central regions (SDSS),
  are the  same at different galactocentric  distances.  Although this
  assumption  is   not  necessarily  true,   the  terms  $C_{\alpha,i}
  \cdot$[$\alpha$/Fe]  vanish  outwards  in  our sample  of  ETGs  (as
       [$\alpha$/Fe]$\sim$0 at  the largest radii  probed).  Thus, the
       computation of $C_{\alpha,i}$ does not affect significantly our
       conclusions.    The  correction  for   non-solar  [$\alpha$/Fe]
       abundance ratios is negligible for TiO features, while it tends
       to increase the  EWs of Mg\,4780 and Ca$_2$,  especially in the
       centre of  the massive  galaxies, where [$\alpha$/Fe]  is high.
       
       In addition, to  allow  for  residual abundance  variations,  we  also
       include the  effect of varying  [Ti/Fe] abundance, as  an extra
       fitting parameter, in Eq.~1 (see LB13 for details).  The reason
       for including [Ti/Fe], rather than any other single element, in
       the analysis is that according to different theoretical stellar
       population models \citep{conroy,  johansson12}, titanium is the
       main  element  whose  individual  abundance ratio  affects  the
       strength of TiO features, and,  in fact, an increase of [Ti/Fe]
       with (central)  velocity dispersion  has been detected  in ETGs
       \citep{johansson12, labarbera}.  Notice that TiO features might
       also  be  affected  by  [O/Fe] abundance  \citep{cgvd}.   Since
       oxygen closely follows  magnesium, with deviations smaller than
       $0.1$\,dex, in  both our Galaxy \citep[independent  of metallicity as
       shown   by][]{Bensby2004}   and  early-type   galaxies
       \citep{johansson12}, the effect  of [O/Fe] is already accounted
       for  by  our  solar-scale  proxy correction  (see  above).
       Moreover,   at    least   some   stellar    population   models
       \citep{johansson12}  do not  predict any  dependence  of either
       TiO$_1$  or  TiO$_2$  on   [O/Fe].   Notice  that  the  [Ti/Fe]
       abundance term  in Eq.~1 does only affect  TiO$_1$ and TiO$_2$,
       while  it is  negligible for  Ca$_2$ and  Mg\,4780.   For what
       concerns  Ca$_2$, this  feature might  also be  affected  by Ca
       abundance variations.   In contrast to Ti,  no {\it residual}
       variation   (i.e.   not  accounted   for  by   our  solar-scale
       correction)  with   (central)  velocity  dispersion   has  been
       detected with Ca.  Therefore, we have not included Ca abundance
       in  the analysis.   The  Mg\,4780 index  could  be affected  by
       radial gradients of several  chemical species (e.g.  C and Si),
       that are difficult to analyse even with the latest state-of-art
       stellar population models.  For  simplicity, we do not consider
       any   of   these  effects   here.    

       Finally,   we  note   that
       gravity-sensitive features are  also sensitive to the effective
       temperature  of RGB stars,  $T_{\rm eff}$.   Decreasing $T_{\rm
         eff}$,  at   optical  wavelengths,  mimics   the  effect  of
       steepening the IMF, but  this degeneracy breaks down completely
       in   the    red   part    ($>   7500$~K)   of    the   spectrum
       \citep{Spiniello2013}.   The  fact that  both  optical and  NIR
       features (e.g.   Na{\sc I}\,8190) suggest an 
       IMF radial  gradient in our massive  galaxies (see Appendix~B),
       excludes  $T_{\rm  eff}$ as  the  main  source  for the  radial
       variation of TiO  features.  Also, one should note  that two of
       the  main sources  of possible  variations of  the $T_{\rm
         eff}$ of the RGB,   i.e.,    Mg   abundance   and    total   metallicity
       \citep{Vandenberg2012}  (besides  Si  abundance),  are  already
       accounted  for in  the  analysis by  means  of the  solar-scale
       correction  procedure  (that   virtually  removes  all  effects
       tightly related  to Mg abundance) and because  we adopt stellar
       population models with varying total metallicity\looseness-2.

\section{Results}

We start discussing below, in a qualitative manner, the radial
behaviour of the IMF-sensitive TiO2 feature (Section~4.1). In
Section~4.2, we perform a quantitative comparison of best-fitting and
observed line-strengths, for all selected IMF-sensitive features,
following the approach described in Section~3.2. The main results of
the present work are presented in Section~4.3, where we contrast the
radial trends of IMF slope for the low-mass galaxy, NGC~4387, and our
best high-mass candidate, NGC~4552. {  Results for our additional
  high-mass galaxy, NGC\,5557, are presented separately in
  Appendix~\ref{sec:NGC5557}, as this galaxy has been analyzed with a
  different methodology than NGC\,4387 and
  NGC\,4552. Appendix~\ref{app:A} also shows that the NIR Na{\sc }8190
  doublet, as well as other optical features, provide (independent)
  radial constraints to the IMF, that are fully consistent with those
  from our fiducial set of spectral features. }

\subsection{Line-strength gradients: qualitative analysis}
\begin{figure}
\begin{center}
\includegraphics[width=8.5cm]{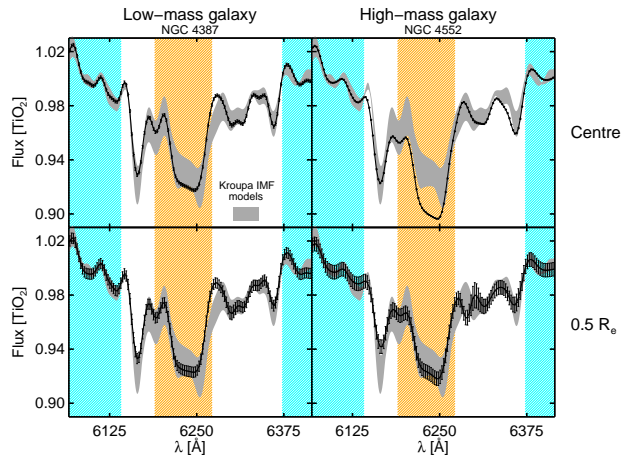}
\end{center}
\caption{Spectral region surrounding  the TiO$_2$ absorption band. The
  Figure  compares the  spectra (black  curves with  1\,$\sigma$ error
  bars) for the low-mass (left) and the high-mass (right) galaxies, in
  their centre (top) and at half of the effective radius (bottom). The
  grey region shows how the TiO$_2$ feature changes for Simple Stellar
  Population (SSP)  models with  fixed, standard (Milky-Way  like) IMF
  over a wide range of  ages and chemical compositions.  Note that any
  linear superposition of SSPs will also fall within the grey regions.
  Observed  and model  spectra  are normalised  to  the blue  flanking
  regions  of  the  (orange-hatched)   TiO$_2$  band,  where  the  IMF
  information is encoded.   In the low-mass galaxy the  TiO$_2$ can be
  well  fitted with a  universal IMF  (grey region)  at all  radii. In
  contrast,  the  massive  galaxy  shows  a  significant  gradient  of
  TiO$_2$: a stronger absorption  is detected in the centre, revealing
  an  enhanced  dwarf-to-giant  ratio.   At  R$\sim  0.5$\,\re,  TiO$_2$
  absorption  is instead  consistent  with the  range  expected for  a
  standard IMF.}
\label{img:qua}
\end{figure}

In  Fig.~\ref{img:qua} the  spectral region  around  the IMF-sensitive
TiO$_2$  spectral feature \citep{trager}  is shown, for two  radial bins
corresponding to the centre (top) and one half of the effective radius
(0.5\,\re, bottom) for NGC~4387  (left) and NGC~4552 (right), respectively.  The TiO$_2$
index measures the absorption strength of a TiO molecular band, prominent in
the atmospheres  of low-mass  cool stars \citep{tio2},  and therefore,
when  detected  in  the  integrated  light of  a  stellar  population,
provides a sensitive tracer of the  ratio of dwarf-to-giant stars. This
ratio is  a surrogate  of  IMF slope:  the higher  the ratio, the steeper the IMF.

The grey shaded regions  in Fig.~\ref{img:qua} show the allowed range
in  the TiO$_2$  feature  for a  wide  range of  age
($7$--$14$\,Gyr),  metallicity ($-0.4$  to  $+0.2$\,dex) and  titanium
abundance  (from $-0.5$ to  $+0.2$\,dex), encompassing  those measured
from our  data, but  keeping the IMF  fixed at the  Milky-Way ``standard''
distribution. We find that: (i)  the TiO$_2$ spectral region of the low-mass
galaxy is well reproduced by Kroupa-like IMF models at all radii; (ii) on
the contrary, for our reference high-mass galaxy, a significant TiO$_2$ mismatch
towards  a   bottom-heavy  IMF  is   detected  only  in   the  central
region. This TiO$_2$ gradient can only be explained by a change of IMF
slope with radius (see  \S\ref{sec:can}).


\subsection{Best-fits to IMF-sensitive spectral indices}\label{sec:grid}

\begin{figure*}
\begin{center}
\includegraphics[width=15cm]{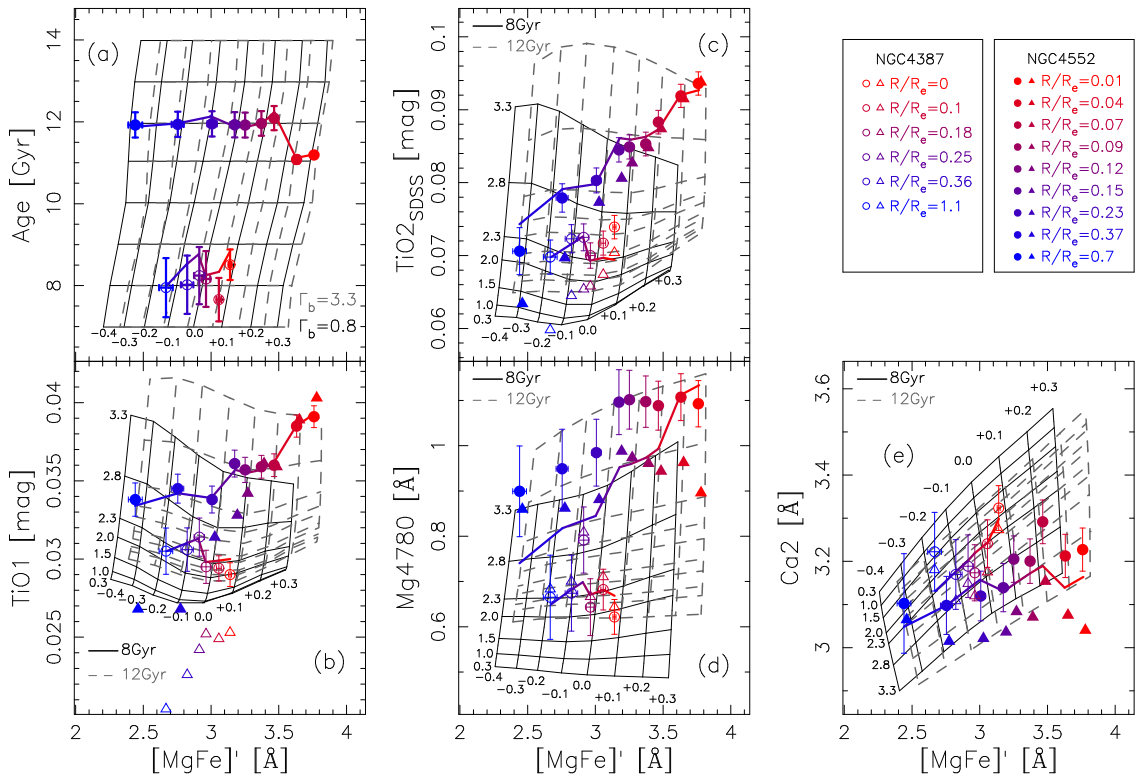}
\end{center}
\caption{   Fitting results for our low- (NGC\,4387; open symbols)
  and high-mass (NGC\,4552; filles symbols) galaxies.  Panel a:
  coloured symbols, with error bars, are age estimates from spectral
  fitting. Notice that the age estimates depend on the assumed IMF
  (see the text). Hence, for each galaxy, and each radial bin, the
  plot shows the age estimate for the corresponding, best-fitting,
  IMF. The grids show the effect of varying metallicity on the $\rm
  [MgFe]'$ index, for SSPs with different ages, and two extreme IMFs
  (grey-dashed and solid-black grids, as labeled).  Panels (b-e):
  IMF-sensitive line-strengths vs. the total metallicity indicator
  [MgFe]$^\prime$.  Triangles show the raw line strengths, whereas
  circles (with error bars) show the line strengths corrected for
  [$\alpha$/Fe], and with the best-fit residual TiO-based abundance
  term ($\delta$Ti in Eq.~1) subtracted off.  The solid-coloured
  curves show the best-fit solutions, with colour varying from red, in
  the galaxy centre, through blue, at larger galactocentric distances
  (see legend in the upper--right corner of the plot).  Solid and dashed
  grids plot line strengths for SSP models with varying
  $\Gamma_\mathrm{b}$ and $\rm [Z/H]$, for two ages of $8$
  (representative of NGC\,4387; see panel a) and $12$~Gyr (oldest
  representative age for NGC\,4552), respectively.  }
\label{fig:grdi}
\end{figure*}

Fig.~\ref{fig:grdi}   illustrates  the   spectral  analysis   for  our
reference low-  and high-mass galaxies, NGC\,4387  (empty symbols) and
NGC\,4552  (solid symbols),  respectively.  Symbol  colours  vary from
red,  in the innermost,  to blue,  in the  outermost radial  bins. The
triangles mark the measured  equivalent widths, while circles plot the
quantities $[ EW_i-C_{\alpha,i}  \cdot [\alpha/{\rm Fe}] - \Delta_{\rm
    Ti,i} \cdot  \delta {\rm  Ti}]$ in Eq.~1,  i.e.  the  observed EWs
corrected  for [$\alpha$/Fe],  and  with the  TiO-based residual  term
removed.  As mentioned above,  the correction in [$\alpha$/Fe] affects
only  Mg\,4780  and  Ca$_2$,   and  is  negligible  for  TiO  features
{($\sim$0.005\,mag  for   the  bin  with   the  largest  [$\alpha$/Fe]
  correction)}. $\Delta_{\rm  Ti,i}$ is  different from zero  only for
TiO-based indices, and in particular for TiO$_1$ (when compared to its
radial gradient; see the large differences among triangles and circles
in panel b).

For  NGC\,4552, the best-fitting  $\delta {\rm  Ti}$ shows  an overall
gradient of about $-0.7$\,dex.   Interestingly, this is similar to the
total metallicity gradient  of this galaxy.  Note that  despite of the
large $\delta$Ti radial variation,  the gradient of TiO$_2$, after the
residual abundance  term is subtracted off, is  still very significant
(filled circles  in panel c),  and can only  be explained by  a radial
gradient of IMF slope. In fact, to  mimic the radial variation of TiO$_2$ with
a  fixed IMF,  we would  require a  large radial  gradient,  more than
$+2$\,dex in  $\delta$Ti.  Such a strong  gradient (at fixed
IMF)  would  be inconsistent  with  the  observed  radial gradient  of
TiO$_1$ (see details in \S\ref{sec:can}).

For Ca$_2$, our best-fitting solutions (Fig.~\ref{fig:grdi}) are fully
consistent (1\,$\sigma$) with the data  in each radial bin.  Note that
this index  decreases with  metallicity, and thus,  because elliptical
galaxies  have   negative  metallicity  gradients  (see   panel  a  of
Fig.~\ref{fig:grdi}), one  would expect a negative  radial gradient of
Ca$_2$.   Apart from  the outermost  radial bin,  where the  error bar
reflects  a large  uncertainty  in sky  subtraction,  such a  negative
gradient  is  clearly  observed  for  the  low-mass  galaxy.   On  the
contrary,  a flatter  trend  is  found in  NGC\,4552,  despite of  its
stronger metallicity gradient.  Considering that Ca$_2$ decreases with
$\Gamma_\mathrm{b}$  \citep{vaz03}, the mild  Ca$_2$ gradient  for the
high-mass galaxy is fully consistent with the result that the slope of its
IMF {  tends}
to  decrease outwards.  The  case of  Mg\,4780 is  less clear,  as this
index still shows a  radial gradient in NGC\,4552, possibly consistent
with a less bottom-heavy IMF outwards.  The gradient is mild, and only
marginally  consistent with  the  other  indices, as  one  can see  by
comparing  the  filled  circles  and  the  best-fitting  solution  for
NGC\,4552 (upper curve) in panel (d) of Fig.~\ref{fig:grdi}.  Based on
stellar population models with varying abundance ratio \citep{conroy},
we  expect  that  this  index anti-correlates  with  several  chemical
abundances, like  [C/Fe] and [Si/Fe]. Hence,  negative gradients, like
those observed for  [Z/H] and $\delta{\rm Ti}$, might  be pushing this
index towards higher values  at large galactocentric distances.  Since
the effect is mild, and we still observe a gradient in Mg\,4780, {  
for the purpose of the present work, we do not attempt any further analysis}.

\subsection{IMF radial gradients}

\begin{figure}
\begin{center}
\includegraphics[width=8.5cm]{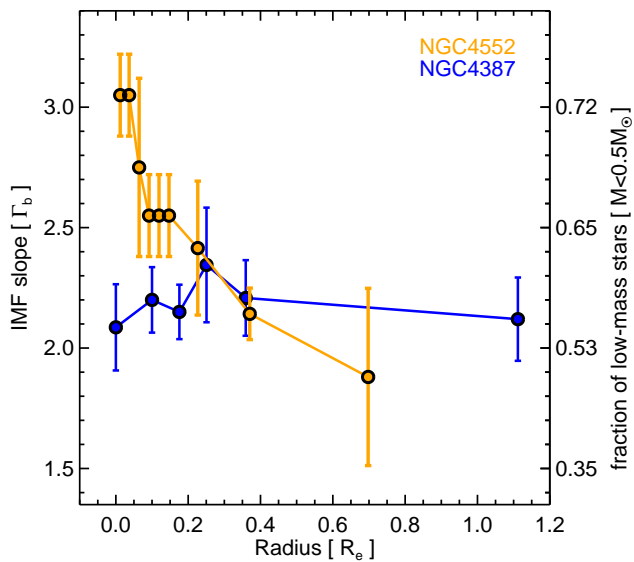}
\end{center}
\caption{  Radial IMF slope profiles for the low- and high-mass
  galaxies, NGC\,4387 and NGC\,4552, respectively.  The IMF slope,
  $\Gamma_\mathrm{b}$, is inferred via a detailed analysis of
  gravity-sensitive features in the galaxy spectra, at different
  galactocentric distances.  The fraction of low-mass stars
  (M$<$0.5M$_\odot$) with respect to the total stellar mass is shown
  in the vertical axis on the right.  The massive ETG (yellow) shows a
  significant IMF slope variation with radius.  The less-massive
  system, NGC\,4387, shows a rather flat IMF radial profile.  Our
  measurements reveal that the enhanced population of dwarf stars
  (i.e.  a higher $\Gamma_\mathrm{b}$) in massive galaxies is confined
  to the central regions.}
\label{img:gamma}
\end{figure}

The main result of this paper is shown in Fig.~\ref{img:gamma}, where
the radial trends of bimodal IMF slope, $\Gamma_b$, are presented for
both NGC~4552 and NGC~4387, respectively.  

  For the massive galaxy NGC\,4552, the IMF slope decreases, by
  $\Delta\Gamma_\mathrm{b}=-1.5$, from the centre to 0.7\,\re.  In the
  innermost radial bins, the inferred IMF slope ($\Gamma_{\rm b}=3.05 \pm 0.2$),
  implies a bottom-heavy distribution, consistent with previous work.
  In constrast, for the outermost bin, the best-fit
  ($\Gamma_\mathrm{b}=1.9 \pm 0.4$) is higher, but still consistent
  at the $1.5\sigma$ level, with the value corresponding to a
  Kroupa-like distribution ($\Gamma_{\rm b} \sim 1.35$).  As shown in
  Appendix~\ref{sec:NGC5557}, the additional massive galaxy
  (NGC\,5557) also exhibits a decrement of $\Gamma_\mathrm{b}$ between
  the central part and the outer region, with $\Delta\Gamma_\mathrm{b}=-1.5 (-0.8)$,
  from the centre to 1(0.7)\,\re.
In contrast, the low-mass ETG, NGC\,4387, shows almost no radial
dependence of IMF within the uncertainties, with
${\Delta\Gamma_\mathrm{b}=-0.3}$.  In Appendix~\ref{sec:corre}, we
also show that the correlation of uncertainties between the best-fitting
parameters (mainly, age, $\delta{\rm Ti}$, and IMF) cannot explain, by
itself, the radial IMF gradient detected in NGC\,4552, i.e. the radial
trend in Fig.~\ref{img:gamma} is not just the result of marginalizing
the uncertainties along a given direction of the parameter space. We
also notice that the inferred IMF slope for NGC\,4387 is higher than
that expected for its velocity dispersion (see LB13).  
 However, while absolute values of $\Gamma_\mathrm{b}$ are
  significantly dependent on the set of adopted spectral features in
  the analysis \citep{Spiniello2013}, it is expected that IMF slope
  variations (i.e.  the radial gradients) are constrained in a much
  more robust way. In addition, one may notice that (i) for a bimodal IMF, the
  difference of IMF normalization (i.e. the stellar mass-to-light
  ratio) between $\Gamma_\mathrm{b}=1.35$ (i.e. a Kroupa-like
  distribution) and $\Gamma_\mathrm{b} \sim 2$ (as we measure for
  NGC\,4387) is only 0.1\,dex \citep[SDSS $r$-band,][]{ferreras}, and (ii) 
  for a Kroupa-like slope ($\Gamma_\mathrm{b}=1.35$), IMF-sensitive features are much 
  less sensitive to a given variation of  $\Gamma_\mathrm{b}$, than for a bottom-heavy
  IMF (e.g.~$\Gamma_\mathrm{b} \sim 3$). Hence, our
  results for the low-mass galaxy do not contradict independent
  constraints from dynamical studies (e.g. the results from the ATLAS3D
  survey), and the mild difference between $\Gamma_\mathrm{b} \sim 2$ and $\Gamma_\mathrm{b} \sim 1.35$ 
  (Kroupa-like) is likely not relevant for the present work.
In terms of the dwarf-to-giant stellar ratio in the IMF (see LB13), up to
$\sim$75\% of the stellar mass in the centre of the massive
galaxy NGC~4552 is accounted for by stars below M$<$0.5M$_\odot$, whereas
for the outermost radial bin, this ratio decreases down to $\sim 50
\%$. For the low-mass galaxy, NGC~4387, stars with M$<$0.5M$_\odot$
account for a roughly constant 56$\%$ of the total stellar mass at all
radii.

{The results of our analysis for 
NGC~4552 and NGC~4387 are summarized in Table~\ref{tab:fit}, where we 
list the best-fitting age, [Z/H], and $\Gamma_\mathrm{b}$
for all radial distance bins, along with the typical signal-to-noise ratio of each 
spectrum. The table also lists the [$\alpha$/Fe] as inferred from the $\rm 
[Z_{Mg}/Z_{Fe}]$ proxy (see Sec.~\ref{sec:proxi}).}

\begin{table*}
\centering
\begin{center}\begin{tabular}{l c c c c c l c c c c c }
\hline
\hline
Galaxy & \multicolumn{1}{c}{$R$} & \multicolumn{1}{c}{$S/N$} & 
\multicolumn{1}{c}{Age} & 
\multicolumn{1}{c}{[Z/H]} & \multicolumn{1}{c}{[$\alpha$/Fe]} 
& \multicolumn{1}{c}{ $\Gamma_\mathrm{b}$} \\
       & (R$_\mathrm{e}$) &    &    (Gyr) & (dex) & (dex)  &   \\
\hline 
NGC~4552     &   0.01 & 1450 &  $11.2 \pm 0.1$  &  $+0.29 \pm 0.01$ &  $0.41 
\pm 0.01$ &  $3.05 \pm 0.2$  \\  
             &   0.03 &  911 &  $11.1 \pm 0.1$  &  $+0.21 \pm 0.01$ &  $0.30 
\pm 0.01$ &  $3.05 \pm 0.2$  \\                      
             &   0.06 &  413 &  $12.3 \pm 0.2$  &  $+0.09 \pm 0.01$ &  $0.30 
\pm 0.01$ &  $2.75 \pm 0.2$  \\                   
             &   0.09 &  281 &  $12.0 \pm 0.2$  &  $+0.04 \pm 0.01$ &  $0.29 
\pm 0.02$ &  $2.55 \pm 0.2$  \\                 
             &   0.11 &  166 &  $11.9 \pm 0.2$  &  $-0.02 \pm 0.02$ &  $0.27 
\pm 0.02$ &  $2.55 \pm 0.2$  \\              
             &   0.14 &  122 &  $11.9 \pm 0.2$  &  $-0.06 \pm 0.02$ &  $0.23 
\pm 0.02$ &  $2.55 \pm 0.2$  \\           
             &   0.22 &  94  &  $11.9 \pm 0.2$  &  $-0.14 \pm 0.02$ &  $0.21 
\pm 0.02$ &  $2.4 \pm 0.3$  \\        
             &   0.37 &  95  &  $12.0 \pm 0.3$  &  $-0.25 \pm 0.02$ &  $0.18 
\pm 0.03$ &  $2.1 \pm 0.1$  \\        
             &   0.69 &  98  &  $12.0 \pm 0.3$  &  $-0.40 \pm 0.03$ &  $0.08 
\pm 0.04$ &  $1.9 \pm 0.4$  \\     
             &    &  &    &   &   &   \\
NGC~4387   &   0.00 & 529  & $8.7 \pm 0.3$  &  $+0.06  \pm 0.01$  &  $0.11 \pm 
0.01$  &  $2.1 \pm 0.2$ \\ 
           &   0.10 & 349  & $7.5 \pm 0.6$  &  $+0.05  \pm 0.02$  &  $0.14 \pm 
0.01$  &  $2.2 \pm 0.2$ \\
           &   0.17 & 187  & $8.0 \pm 0.6$  &  $-0.02 \pm 0.02$  &  $0.12 \pm 
0.02$  &  $2.1 \pm 0.1$ \\
           &   0.25 & 143  & $8.1 \pm 0.6$  &  $-0.05 \pm 0.03$  &  $0.09 \pm 
0.02$  &  $2.3 \pm 0.3$ \\
           &   0.35 & 110  & $7.9 \pm 0.7$  &  $-0.09 \pm 0.03$  &  $0.13 \pm 
0.02$  &  $2.2 \pm 0.2$ \\
           &   1.11 & 93   & $7.8 \pm 0.6$  &  $-0.16 \pm 0.03$  &  $0.09 \pm 
0.03$  &  $2.1 \pm 0.2$ \\   
\hline 
\end{tabular}
\caption{  Best-fitting parameters for the massive galaxy NGC~4552,
  and the low-mass galaxy NGC~4387. Uncertainties are quoted at the
  one sigma level.  }
\label{tab:fit}
\end{center}
\end{table*}

\section{Robustness of inferred IMF gradients}\label{sec:test}

To prove the robustness of our results, we have performed a battery of
tests, varying the procedure to infer the IMF slope.  These tests are
described below.

\begin{description}
\item[\textbf{Age estimate:}] This is one of the main uncertainties in
  the IMF determination, as most IMF-sensitive spectral indices are
  sensitive to age. In addition to including the Ca2 index in the
  analysis (which is sensitive to IMF, but not to age), we have
  addressed the age estimate issue as follows. (i) We have tested the
  effect of more complex star formation histories (than a single SSP),
  estimating the age from spectral fitting with two-SSP models.  (ii)
  As a further {\it extreme} test, we have also calculated the radial
  IMF profiles by neglecting completely the age constraint from
  spectral fitting, i.e.  repeating the fits by removing the first
  term in the right-hand side of Eq.~1.  (iii) We have tested the
  impact of our assumption that residual $\delta{\rm Ti}$ correction
  affect only spectral indices, and not the age determination
  itself. To this aim, we have implemented an iterative approach,
  where spectral and index fitting are repeated, after dividing each
  input spectrum with the ratio of [Ti/Fe]-enhanced to solar-scale SSP
  theoretical models \citep{conroy}, the ratio being scaled to match
  the [Ti/Fe] residual abundance from the first minimisation of Eq.~1.
  In practise, the output best-fitting parameters from the first
  iteration are very similar (within a few percent) to those from the
  zero-order step, requiring no further iterations.

\item[ \textbf{Temperature effects on the SSP modelling:}] We have
  varied (iv) the temperature scale of low-mass
  ($M$\,$<$\,0.5\,$M_\odot$) dwarf stars in MILES models -- which has
  a major impact on various IMF-sensitive indicators -- according to a
  cooler prescription \citep{ss} than the MILES reference one
  \citep{miuscat}; (v) we have arbitrarily shifted the scale of RGB
  stars by $\pm 50$~K, as these changes affect significantly molecular
  bands (i.e. TiO features).

\item[ \textbf{Flux  calibration uncertainties:} ] We  have fitted the
  continuum of  each observed spectrum,  in the TiO  spectral regions,
  with MILES SSPs multiplied by  a low-order polynomial in each region
  (with degree  $\delta(\lambda)/100$, where $\delta(\lambda)$  is the
  spectral range size,  see \citet{conroy}).  Fitting different models
  (e.g.  varying slightly  the IMF,  age, metallicity  of  MILES SSPs,
  and/or  the  polynomial  degree,   and  for  different  spectra)  we
  estimated  that the  impact  of flux  calibration uncertainties  may
  affect significantly TiO$_1$, and to less extent TiO$_2$.  Even accounting
  for  rotation,  the  net  effect  is to  offset  the  observed  line
  strengths (i.e.   radial gradients are not affected  at all).  Thus,
  as a  further test, (vi) we  have repeated the  analysis by shifting
  TiO$_1$ and  TiO$_2$ line-strengths by  the maximum shifts allowed  by our
  continuum fits, i.e.   $+ 0.009$~mag and $+0.005$~mag, respectively.
  The outcome  of this test is  that TiO flux  calibration offsets can
  introduce  a large  shift in  the  derived $\delta{\rm  Ti}$, up  to
  $\sim$0.6~dex, while IMF gradients remain unchanged. Hence, the absolute
  value of the $\delta{\rm  Ti}$ does not reflect the actual [Ti/Fe]
  overabundance of the stellar population.

\end{description}

\begin{figure}
\begin{center}
\includegraphics[width=8.5cm]{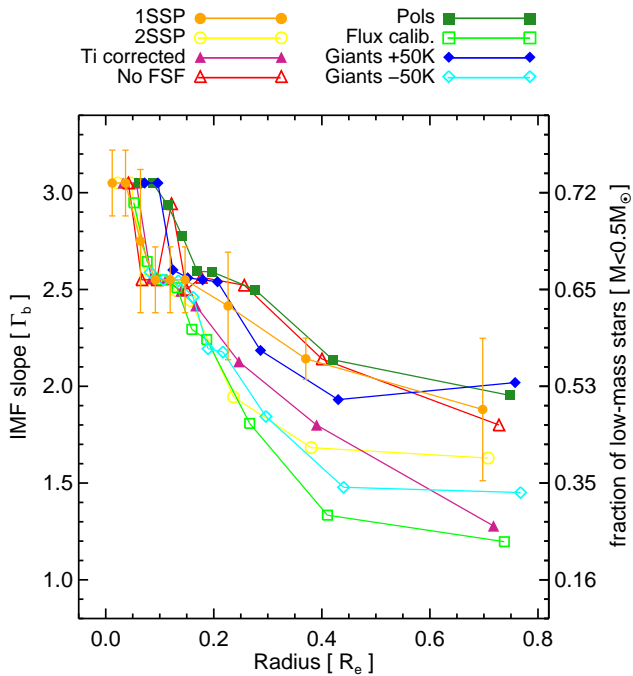}
\end{center}
\caption{ Different IMF radial profiles {for NGC~4552}, obtained by
  changing several assumptions in the modelling/fitting process.
    Profiles obtained with different assumptions are plotted with
    different colours and symbols, as labeled on the top of the
    figure.  Remarkably, all radial profiles show a gradient in the
    IMF slope, from a bottom-heavy IMF in the galaxy centre to a more
    ``standard'' slope, closer to a Kroupa-like value of $\rm
    \Gamma_{b}=1.3$, at about one effective radius. Hence, our
  results are robust against all performed tests: age derived through
  one (\textit{1SSP}) or two (\textit{2SSP}) SSPs; considering the
  $\delta{\rm Ti}$ impact on the age estimation (\textit{Ti
    corrected}); inferring the age with no information from full
  spectral fitting (\textit{No FSF}); using a cooler prescription for
  the dwarf stars temperature (Pols); testing the flux calibration
  (\textit{Flux~calib.}) and changing the temperature scale of giant
  stars by plus 50 K (\textit{Giants $+50$K}) and minus 50 K
  (\textit{Giants $-50$K}).}
%
\label{fig:tests}
\end{figure}

The results of the above tests (i--vi) are shown, -- for our reference massive galaxy,
NGC\,4552 --  in  Fig.~\ref{fig:tests}.  The Figure plots IMF radial  trends
for all tests,  proving that indeed our results are very robust. 
In particular, none of our experiments -- changing the assumptions
of fitting methodology, stellar population modelling,  and
exploring  the impact of  flux calibration  issues in  the data  -- 
alters the existence of a significant radial IMF gradient for NGC\,4552.
 
\section{Can we fit the observations with  a constant IMF?}\label{sec:can}

In the following, we elaborate on the tests presented in Section~\ref{sec:test},
discussing why a  radial variation of the IMF seems to be the  only possible  
explanation to the  radial trends of IMF-sensitive indices  in
our massive galaxies. To this effect, we discuss several  scenarios 
that might mimic the  effect of a varying IMF with galacto-centric distance:

\begin{enumerate}
\item \textbf{Varying the temperature scale of dwarf ($<$0.5\msun) stars.}
As mentioned in test (iv) of Section~5,   
{we  have   constructed  SSP   models  with  a   significantly  cooler
  prescription     \citep{ss}      for     low-mass     dwarves than MILES   SSPs
  \citep{miuscat}, by  $\Delta$T$\sim$\,200\,K.  This  temperature variation  
  turns  out to  affect
  mostly the  TiO indices.  For  an IMF slope of  $\Gamma_\mathrm{b} =
  3$, the TiO$_2$ decreases by $\sim 0.007$ mag (when using the cooler
  temperature scale).  For $\Gamma_\mathrm{b} = 1.35$ (i.e. Kroupa-like IMF), the TiO$_2$ 
  changes  only by $\sim  -$0.002 mag.  Since the  metallicity  proxy, $\rm
  [MgFe]'$, is almost insensitive to the dwarf temperature scale, the
  net effect  is to  shrink the model  TiO$_2$--$\rm [MgFe]'$  grid in
  panel   c   of   Fig.~\ref{fig:grdi},   along   the   ``IMF   axis''
  (y-axis).  Hence, although  a  change in  the  temperature scale  of
  low-mass dwarves   affects the absolute values of  the IMF slope
  in the galaxy centre, the  IMF gradient cannot be removed, as shown
  in Fig.~\ref{fig:tests}. }

\item \textbf{Varying the temperature scale of giant stars.} 

We  have constructed  SSP models  by shifting  the temperature  of RGB
stars by $\delta(T_{\rm eff})=-150$, $-100$, $-50$, $0$, and $+50$\,K,
respectively  (a shift of  zero corresponding  to our  reference MILES
models).   For  $\delta(T_{\rm   eff})=-150$\,K,  the   model  TiO$_2$
increases by  $\sim$0.02\,mag, i.e.  a large  variation, comparable to
that  between a  bottom-heavy  and standard  IMFs.   A similar  result
applies  to Mg\,4780  and  TiO$_1$. Therefore,  a  radial gradient  of
$T_{\rm  eff}$ might be  able to  explain, by  itself, the  TiO radial
gradients. {    However, this possibility is ruled out by our NIR spectral
indices, as well as other gravity-sensitive features (CaH1 and CaH2, see App.~\ref{app:A}).  
In particular, for $\delta(T_{\rm  eff})=-150$~K, the  model  Ca2  index
changes by  only $\sim$0.03\,\AA, while the model  Na{\sc I}\,8190 shows
an opposite  variation} with respect  to its observed  radial gradient,
{\it decreasing} {  by $\sim $0.2\,\AA }.   We notice that two of the main
sources of  opacity in a stellar  population  --
total metallicity  and [Mg/Fe], besides [Si/Fe] \citep{Vandenberg2012} -- 
are  properly accounted for  in our  analysis, the
former by  the  fact  that  we  rely on  models  with  varying  total
metallicity, and the latter by our [$\alpha$/Fe] correction procedure.
{  This leaves very small leeway for residual $T_{\rm eff}$
variations.  
{\it    Furthermore, as already mentioned, a positive radial  gradient of $T_{\rm
    eff}$ cannot  match simultaneously  the observed radial  trends of
  both  TiO, Na, and Ca/CaH indices.   }  }  Since  a  change of
$T_{\rm  eff}$ also  affects  the shape  of  the TiO-  [MgFe]$^\prime$
index-index grid  (see Fig.~\ref{fig:grdi}), we have also  tested the
impact of a fixed  $\delta(T_{\rm eff})$ ($=+/-50$\,K) on our results.
Since the change  in the shape of  the TiO1 vs. [MgFe]$^\prime$  and TiO2 vs
[MgFe]$^\prime$ grids  is very similar, no significant
variation is found in the IMF trends (see Fig.~\ref{fig:tests}).

\item \textbf{Radial [$\alpha$/Fe] variation.}

As found in LB13, the [$\alpha$/Fe]  correction of TiO indices is small
(e.g.  less  than  $\sim$0.005\,mag for  TiO$_2$).  Therefore,
radial gradients of [Mg/Fe] (and  that of other $\alpha$ elements that
closely  follow  Mg, such as [O/Fe]) can  not  explain  the  radial variation  of  TiO
indices.  Furthermore, the measured  gradient in [$\alpha$/Fe] in both
massive galaxies would imply a  Na{\sc I}\,8190 gradient opposite to the
observed  one (i.e., in this scenario,  Na{\sc  I}8190 should  increase outwards).

\item \textbf{Radial [Ti/Fe] abundance variation.}

A radial gradient of [Ti/Fe], as  large as 2~dex, would be required to account
for  the  observed  gradient of the TiO$_2$  index.
Because of  the expected  similar response of  both TiO indices  to Ti
abundance, one cannot fit  both TiO$_1$ and
TiO$_2$  gradients with  a radial  variation of  [Ti/Fe] alone.  If we
leave only age, metallicity,  and [Ti/Fe] abundance as free parameters,
for models with a fixed IMF, the best-fitting solution for the massive
galaxy NGC\,4552 deviates by  more than 2\,$\sigma$ (3\,$\sigma$) from
the  observed TiO$_1$  (TiO$_2$)  values, as shown in Fig.~\ref{fig:fixIMF}.
Thus, a  radial variation  of [Ti/Fe] cannot mimic the effect of a radially varying IMF.

\begin{figure*}
\begin{center}
\includegraphics[width=12cm]{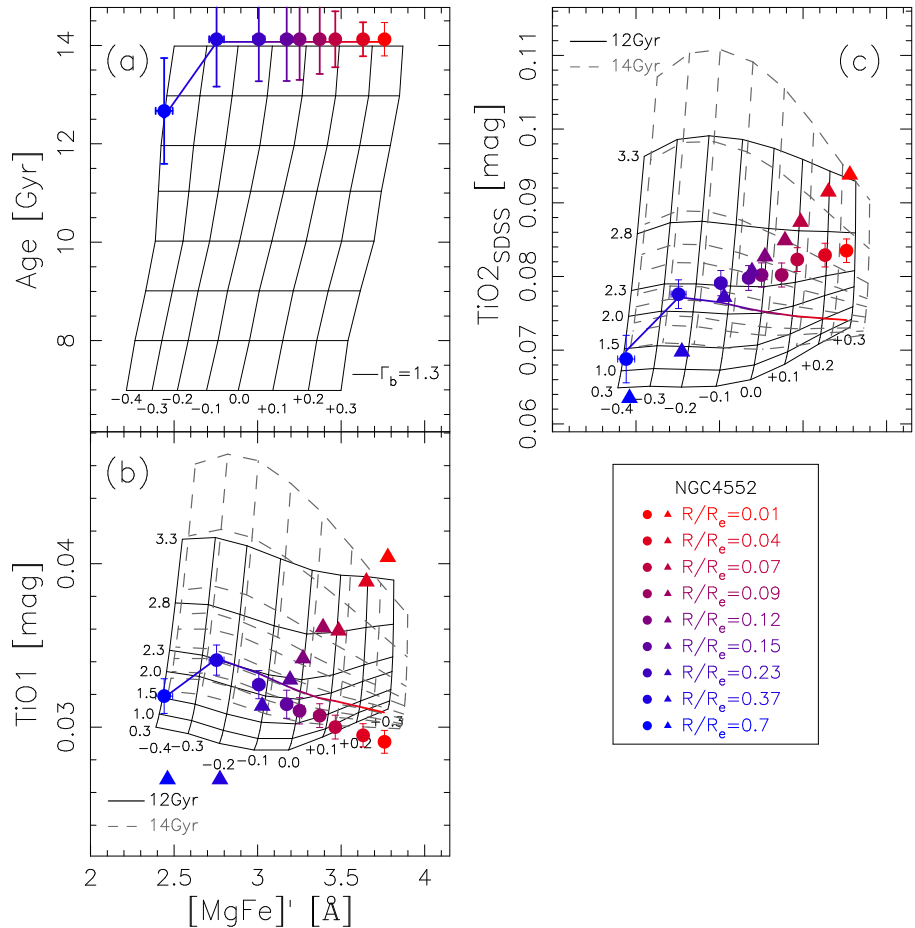}
\end{center}
\caption{  Fitting results for our reference high-mass galaxy,
  NGC\,4552, obtained by leaving age, metallicity, and $\delta$Ti as
  free parameters, but assuming a fixed, standard (Kroupa-like), IMF.
  Panel (a): coloured dots, with error bars, are age estimates from
  spectral fitting, for a Kroupa IMF. The grid shows the effect of
  varying metallicity on the $\rm [MgFe]'$ index, for SSPs with a
  Kroupa-like IMF ($\Gamma_\mathrm{b}=1.35$) and different ages.
  Panels (b-c): solid and dashed grids show line strengths for SSP
  models with varying $\Gamma_\mathrm{b}$ and $\rm [Z/H]$, for two
  representative ages of $12$ (outermost radial bin; see panel a) and
  $14$~Gyr (innermost bins), respectively. Filled circles, with error
  bars, are the [$\alpha$/Fe]-corrected and $\delta$Ti shifted line
  strengths, while triangles plot raw line strengths (as measured on
  the spectra). Notice that triangles are the same as in panels (b-c)
  of Fig.~\ref{fig:grdi}. Since $\delta$Ti is a free fitting
  parameter, the $\delta$Ti values are not the same when assuming a
  fixed, relative to a varying, IMF. Hence, coloured dots in panels
  (b-c) are not the same as those in Fig.~\ref{fig:grdi}.  In all
  panels, solid-coloured curves show the best-fit solutions, with
  colour varying from red, in the galaxy centre, through blue, at
  larger galactocentric distances.  The same colour coding applies to
  dots and triangles (see the legend in the lower--right corner of the
  plot).  For the two outermost bins, the best-fit Kroupa-like models
  (coloured curves) match the [$\alpha$/Fe]-corrected and $\delta$Ti
  shifted line strengths (filled circles).  On the contrary, in the
  five innermost bins, the model TiO$_2$ deviates by more than
  3\,$\sigma$ from the observations, while significant deviations (at
  2\,$\sigma$) are also seen for TiO$_1$. }
\label{fig:fixIMF}
\end{figure*}

\item \textbf{Radial [Na/Fe] abundance variation.}

{  In Appendix~\ref{sec:na}, we analyze the radial behaviour of the
  Na features at $5900$\,\AA \ (NaD) and $8200$\,\AA \ (Na{\sc I}8190),
  respectively.  For a constant IMF, after subtracting off the
  expected variation due to the metallicity gradient of NGC\,4552, one
  could explain the radial variation of Na{\sc I}8190 with a gradient
  of about 1~dex in [Na/Fe] abundance (with an uncertainty of $\pm
  0.2$~dex, depending on the data reduction procedure).  Such a
  scenario would imply a change of about 3.3\,\AA\ in the NaD line
  strength. {  In contrast, a modest change is allowed by the
    observations ($<$1\,\AA , after removing the effect of
    metallicity; see bottom panel of Fig.~\ref{fig:na} and
    Appendix~\ref{sec:na}).}  Hence, similarly to TiO indices, we
  cannot explain multiple indices (in this case, the NaD and Na{\sc
    I}8190 radial gradients) by invoking a variation of [Na/Fe] alone,
  with the assumption of a fixed IMF.  }

\end{enumerate}

{  Finally, we want to emphasize that 
the agreement in the radial behaviour of  a wide set of spectral features (from 
different chemical species and covering a large wavelength range) 
as that considered in this paper (i.e. TiO1, TiO2, Ca2, Mg\,4780 plus aTiO, CaH2, NaD, NaI\,8190) 
breaks the degeneracy between IMF  and elemental abundances and/or $T_{eff}$ (see
  App.~\ref{app:A} and Sec.~\ref{sec:test}), proving the robustness of our result, i.e. the IMF radial variation
  in the two massive ETGs studied here.
  }


\section{Discussion}

  We have found that the massive ETG NGC~4552 shows a
radial  variation  of  gravity-sensitive  indices  that  can  only  be
explained if  an IMF radial  gradient is invoked, from  a bottom-heavy
slope (i.e.,  an enhanced  dwarf-to-giant ratio) in  the centre,  to a significantly flatter 
distribution, closer to the  
standard   (Kroupa-like)  IMF,  at a galacto-centric distance of about one effective
radius. Evidence for a decrement in IMF slope are also found for the additional
high-mass galaxy NGC\,5557, whose IMF slope is fully consistent with a Kroupa-like 
IMF in the outermost radial bin. On the  contrary, a low mass ETG,  NGC~4387, presents a 
flat radial trend of IMF slope.

Hence, although previous studies suggested a trend with galaxy mass on
a   global  sense   \citep{cenarro,vandokkum,ferreras,labarbera},  our
findings imply that {\sl it is in the cores of massive galaxies} where
star  formation processes  are fundamentally  different, leading  to a
bottom-heavy    IMF.      The    $\Gamma_\mathrm{b}$    profiles    of
Fig.~\ref{img:gamma}  imply  that the  correlation  of  IMF slope  and
galaxy               mass               found in previous studies
\citep{cenarro,vandokkum,cappellari,ferreras,labarbera} likely arises because
of  an enhanced  fraction of  low-mass  stars in  the central  regions
(R$\lesssim   0.3-0.5$\,\re  )   of  the   most  massive   ETGs.   The
$\Gamma_\mathrm{b}$-mass  relation \citep{ferreras,labarbera} predicts
$\Gamma_\mathrm{b} \sim 3$ for $\sigma \sim 300$\,\kms , a value fully
consistent  with  the  innermost  radial  bins  of  NGC\,4552.  It  is
therefore  the central  region  that drives  the observed  correlation
between IMF slope and velocity dispersion among galaxies.

Our results suggest that the IMF in ETGs is driven by the local
conditions of the interstellar medium during the process of formation.
At present, there is no complete theory of star formation that could
address in a comprehensive way the connection between the initial
stages of star formation and the properties of the IMF, especially the
characteristic stellar mass and the location of the turnover in the
power law at low masses.  The fragmentation scales are driven by
complex physics involving thermal properties of the gas; the
contribution of dust to cooling; magnetic fields; turbulence; the
effect of supersonic motions in the turbulent gas; and the transition
from pre-stellar cores to stars, among others
\citep{padoan:02,larson:05,bonnell,Hopkins:13}.  In addition,
constraints based on observations of unresolved stellar populations
can only provide an effective IMF integrated with respect to the past
star formation history of the galaxy \citep[i.e.  the integrated
  galactic IMF,][IGIMF]{igimf}.  For instance, a na\"\i ve assumption of a
bottom-heavy IMF in the cores of massive galaxies would lead to a
significant paucity of metals, in conflict with the observed
metal-rich populations in these regions.  It is by assuming that the
IMF is time-dependent, and coupled to the star formation rate, that
one can obtain results that are compatible with the observations
\citep{vazdekis:97,elsca:06,weidner:13}. Nevertheless, it is not clear
at present how the most developed theoretical mechanisms describing
variations in the IGIMF over galactic scales \citep{weidner:05} can
also explain local variations.  Therefore, observational constraints
such as the one presented in this paper provide essential information
towards a complete theory of star formation.

Furthermore,  the observed  radial difference  in the  IMF  of massive
galaxies can be understood as  the imprint from two different modes of
galaxy formation \citep{oser}: a  quick and efficient one, building up
the  cores of massive  ETGs at  high redshift,  during an  early epoch
\citep{mono} -- leading to a  system with high velocity dispersion and
a  bottom-heavy  IMF --  followed  by a  more  gradual  growth of  the
outermost regions \citep{navarro}, perhaps  either by the accretion of
smaller   structures  \citep{naab},   or  via   a   quiet,  prolonged,
star-formation  rate,  where  the  lower velocity  dispersion  of  the
interstellar medium leads to a  standard IMF.  Studies of massive ETGs
with  uncorrelated  radial  gradients  of  the  IMF  distribution  and
[$\alpha$/Fe] enhancement are crucial to determine the local parameter
driving the  IMF in these galaxies. This  two-phase formation scenario
is in sharp  contrast with a na\"\i ve interpretation  of the hierarchical
buildup  paradigm of  galaxy formation,  where high-mass  systems form
through  the mere  assembly of  smaller units,  and calls  for further
observational and theoretical advances.

\section{Conclusions}

We  have  analysed the  spectra  of  two nearby ETGs:  a massive
galaxy, NGC~4552,  and   a   low-mass  counterpart, NGC~4387. 
Results for an additional high-mass galaxy, NGC\,5557 (with lower quality data), 
are also discussed.
Comparing observed line-strengths to predictions of 
state-of-art stellar population models, we have found that:

\begin{enumerate}

 \item Massive galaxies show a steep radial variation of the IMF
   slope, with an enhanced fraction of low-mass stars in the centre
   and a standard Kroupa-like distribution at the effective radius
   (Fig.~\ref{img:gamma}).

 \item The IMF gradient of the low-mass galaxy is rather flat, 
    mildly steeper than that for a Kroupa-like IMF throughout (Fig.~\ref{img:gamma}).

 \item Our result naturally explains the IMF-slope vs central velocity
   dispersion relation of ETGs, as a luminosity-weighted average of
   the underlying IMF radial gradient,
 
\end{enumerate}

We therefore suggest that the IMF of nearby ETGs should be
regarded as a ``local'' property, with an excess of low-mass stars
being produced by processes driving the formation of their cores --
during the early phases of star formation.

Although  simple, phenomenological  models provide  explanations to
this   scenario  \citep{Hopkins:13,weidner:13},  detailed   ab  initio
numerical simulations are required to understand this fundamental 
link between  the growth of  structures and the ``baryon''  physics of
galaxy formation.

\vspace{0.2in}

\small{ \footnotesize{\textit{{Acknowledgements}}} {  We
    would like to thank the anonymous referee, for the many helpful
    comments, that helped us to significantly improve this
    manuscript.}  We would like to thank the GTC astronomers Antonio
  Cabrera, David Garc\' ia and Antonio Garc\'ia for their work during
  the observations. This work has benefited from interesting
  discussions with C. Weidner and M.  Beasley. IMN would like to
  specially thank Luis Peralta de \mbox{Arriba} for his comments
  during this work.  This work has been supported by the Programa
  Nacional de Astronom\'ia y Astrof\'isica of the former Spanish
  Ministry of Science and Innovation under grant AYA2010-21322-C03-02.
  {Based on observations made with the Gran Telescopio Canarias
    (GTC), installed in the Spanish Observatorio del Roque de los
    Muchachos of the Instituto de Astrof\'\i sica de Canarias, in the
    island of La Palma.}}

\bibliographystyle{mn2e}
\bibliography{IMF.bib}

\newpage

\appendix

\begin{center}
\section{NGC~5557} \label{sec:NGC5557}
\end{center}

\begin{figure*}
\begin{center}
\includegraphics[width=13cm]{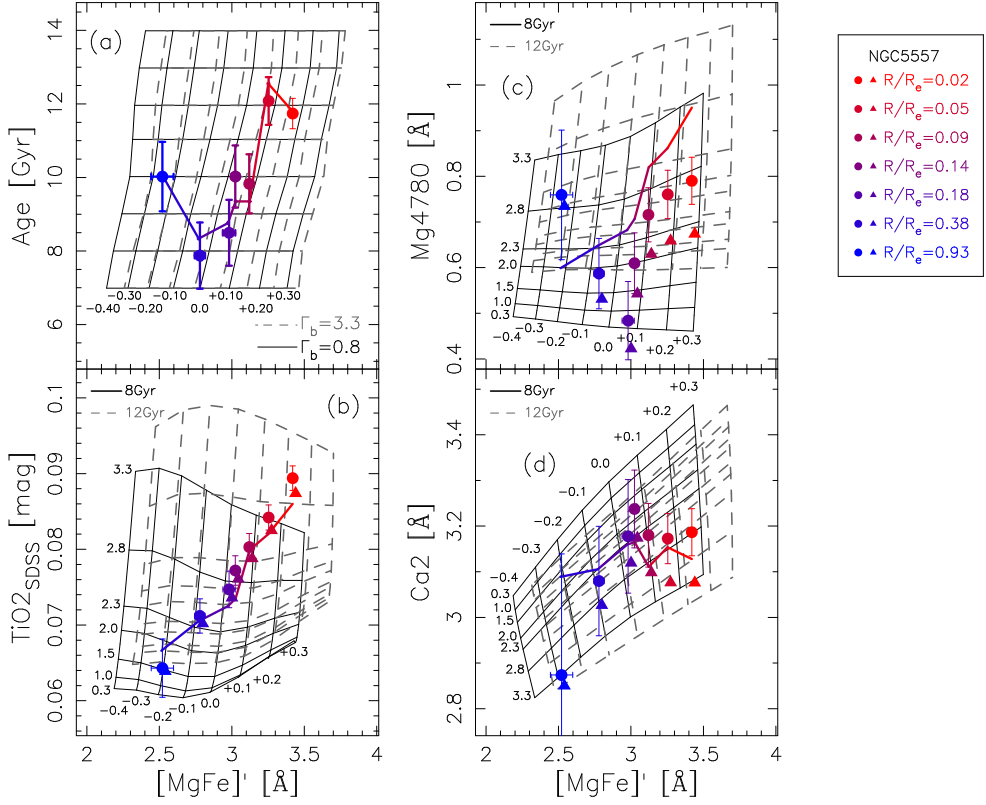}
\end{center}
\caption{  Fitting  of line  strengths  for the  massive ETG  NGC\,5557.
  This  Figure is  similar to  Fig.~\ref{fig:grdi}, but  excluding the
  TiO$_1$  spectral feature,  which is  not reliable  for  this galaxy
  because  of strong contamination from telluric  absorption.
  Notice also  that  the  solar-scale correction  (difference  between
  triangles and circles in panels b--d) does not include the effect of
  $\delta$Ti residual abundance, as for NGC\,4387 and NGC\,4552 in Fig.~\ref{fig:grdi}. }
\label{fig:ngc5557grids}
\end{figure*}

\begin{figure}
\begin{center}
\includegraphics[width=8.5cm]{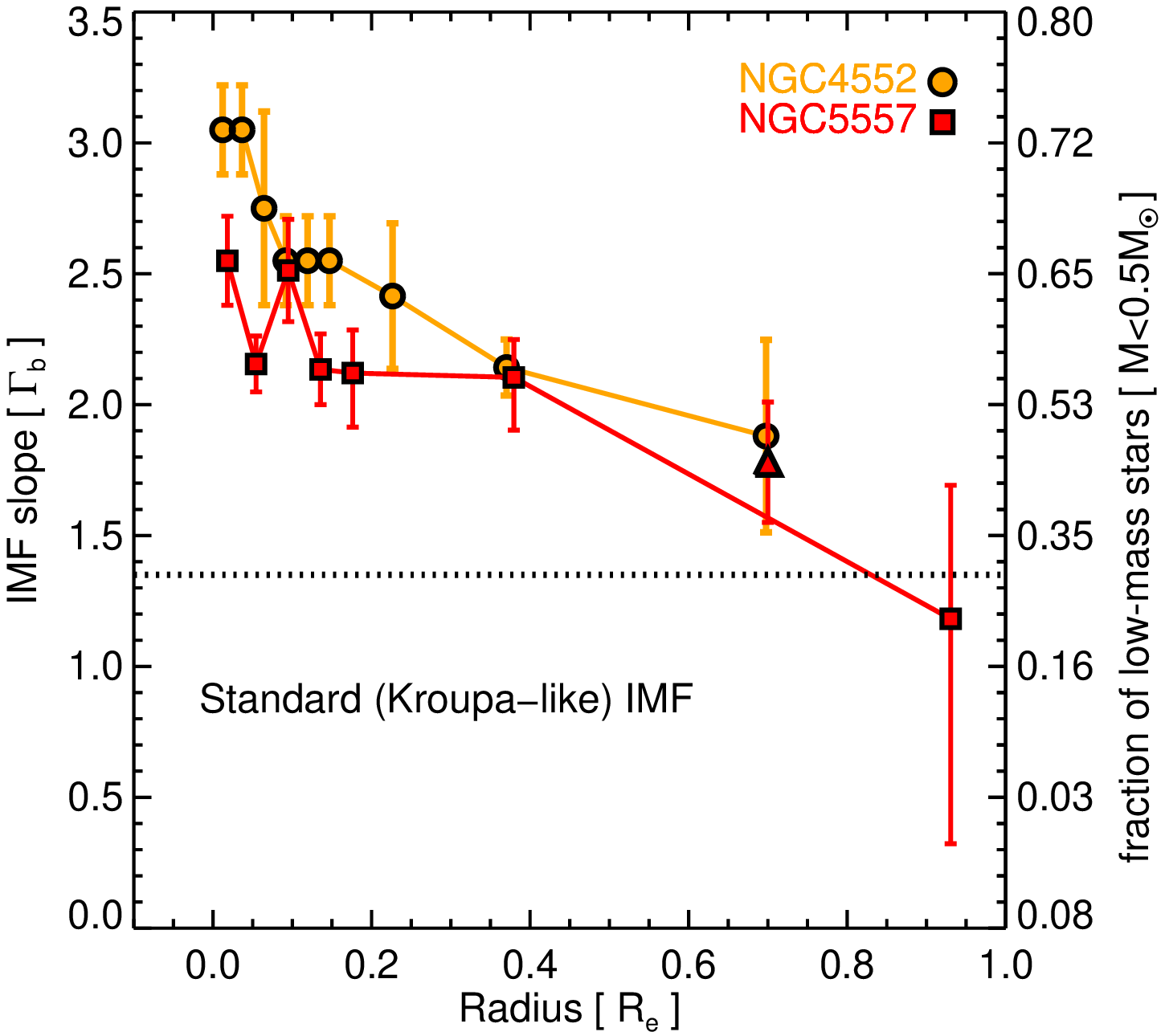}
\end{center}
\caption{  Radial IMF-slope gradient for the massive galaxy
    NGC\,5557 (orange circles), compared to that for our reference
    massive ETG, NGC\,4552 (red squares).  The trend for NGC\,4552 is
    the same as in Fig.~\ref{fig:grdi}. Error bars denote
  1\,$\sigma$ uncertainties.  Notice that error bars are
    asymmetric, reflecting the fact that gravity-sensitive features
    are more sensitive to high, relative to low, values of $\rm
    \Gamma_{b}$. Because of the different fitting methodology (see the
    text), the results for NGC\,5557 are only presented in this
    Appendix. Despite the relatively large error bars, the results
    for NGC\,5557 also indicate a gradient of IMF slope, varying from
    bottom-heavy in the centre to Kroupa-like outwards (see dotted
    horizontal line in the bottom panel), corroborating the result for
    our reference massive galaxy, NGC\,4552.  A radial bin at
    R~=~$0.7 \rm R_{e}$ (red triangle) was calculated to compare directly both NGC\,4552
    and NGC\,5557 IMF gradients at the same radial distance. Although this new bin is not
    independent from the original profile, it shows that the derived IMF
    gradient of NGC\,5557 is not due to a larger error in the last radial bin.}
\label{fig:ngc5557imf}
  \end{figure}

{  As part of the same observing campaign with OSIRIS/10.4m GTC, we
  also acquired deep spectroscopic data for the high-mass ETG
  NGC\,5557.  Similar to NGC\,4552, this galaxy shows no kinematic
  peculiarity within its effective radius, with a velocity dispersion
  profile that declines smoothly from $\sim 300$\,\kms\ in the centre,
  to $\sim 200$\,\kms\ at about one effective radius.  We reduced the
  data for NGC\,5557 in the same way as for NGC\,4552 and NGC\,4387.
  In particular, we extracted spectra in different radial bins, after
  removing the effect of the rotational velocity and velocity
  dispersion at each position along the slit. The radial bins were
  defined adaptively, in order to reach a target $S/N$ \break ratio
  $\geq$ 100 per \AA \ for each bin. 
  
  Unfortunately, due to the recession velocity of the galaxy, the spectrum of 
 NGC\,5557 is redshifted to an observed frame where both TiO$_1$ and (to a minor
  extent) TiO$_2$ are affected by telluric absorption.  We evaluated
  this effect by applying (alternatively) two different telluric
  absorption models to the observed spectra of NGC\,5557, at all
  radial positions (i.e.  each row along the two-dimensional spectrum
  of the galaxy). The two models were constructed in a similar way as
  detailed below for NGC\,4552 and NGC\,4387 (see Sec.~\ref{sec:na},
  where the two models are referred to as $TELL1$ and $TELL2$,
  respectively). The TiO indices were re-measured on the corrected
  spectra, re-extracted in each radial aperture. For both telluric
  models, the correction was found to be relevant for the TiO$_1$
  index, with a variation of up to $\sim$0.025\,mag in the innermost
  radial bin (i.e. comparable to the entire dynamical range covered by
  this index in panel b of Fig.~\ref{fig:grdi}).  On the contrary, the
  effect of telluric absorption on the TiO$_2$ was very small,
  amounting to less than $\sim 0.003$\,mag at all radial bins (i.e.
  about one tenth of the difference between $\Gamma_\mathrm{b}=0.3$
  and $\Gamma_\mathrm{b}=3.3$).  
  
  For these reasons, we have analyzed
  the spectra of NGC\,5557 with a different approach, by excluding
  TiO$_1$ from the $\chi^2$ minimization procedure (see
  Eq.~\ref{eq:method}).  The TiO$_2$ index is also mildly contaminated
  by airglow, affecting only the central part of the feature, from
  6224 to 6248\,\AA.  We interpolated this spectral region both in
  models and observed data, for a consistent analysis.  We point out
  that the effect of the interpolation is completely negligible,
  amounting to less than $\sim 0.002$\,mag on the TiO$_2$
  line-strengths, at all radial bins.
{Since there is  a degeneracy between IMF slope  and [Ti/Fe] abundance
  on a  single TiO feature, for  NGC\,5557 we have  minimised Eq.~1 by
  neglecting the effect of  possible Ti abundance variations.  This is
  motivated by  the fact that, for NGC\,4552,  this residual abundance
  is found to have a minor effect on TiO$_2$ (see panel c of
  Fig.~3),  both NGC~4552  and  NGC~5557  showing  a similar  TiO$_2$
  gradient ($\sim 0.025$~mag).}

Figs.~\ref{fig:ngc5557grids} and~\ref{fig:ngc5557imf} show the fitting
of line-strengths for NGC\,5557 and the resulting gradient of IMF
slope, respectively. 
 The $\Gamma_\mathrm{b}$ radial profile for our reference massive
  galaxy, NGC\,4552 (see Fig.~\ref{img:gamma}), is overplotted in
  Fig.~\ref{fig:ngc5557imf}, in order to allow for a direct comparison of
  the two massive galaxies. As for NGC\,4552, the Mg\,4780 index of
  NGC\,5557 deviates more than the other indices with respect to the
  best-fitting solution, although a radial gradient is detected for
  this feature (see panel c of Fig.~\ref{fig:ngc5557grids}).  Overall,
  the $\Gamma_\mathrm{b}$ radial profile of NGC\,5557 shows a similar
  behaviour as for NGC\,4552, both profiles pointing to a decrease of
  IMF slope with galactocentric distance.  However, one can notice
  that the $\Gamma_\mathrm{b}$ values for NGC\,5557 are systematically
  lower than those for NGC\,4552. This might be due to the fact that
  we have analyzed these two galaxies with different sets of spectral
  features, or, alternatively, NGC\,5557 might have a less
  bottom-heavy IMF than NGC\,4552. For the purpose of the present
  work, the relevant point is that the analysis of NGC\,5557 also
  points to a radial IMF gradient, with the bottom-heavy population
  confined to the galaxy central regions.  We notice that
  Fig.~\ref{fig:ngc5557imf} suggests that the IMF radial gradient for
  NGC\,5557 might be less significant than that for NGC\,4552, mostly
  because of the relatively large error bars on $\Gamma_\mathrm{b}$ in
  the outermost radial bins.  To address this issue, we have fitted
  the $\Gamma_\mathrm{b}$ vs. $\rm R/R_{e}$ trend for NGC\,5557 with a
  linear relation, adopting an ordinary least-squares fitting procedure
  with $\Gamma_\mathrm{b}$ as dependent variable. We have repeated the
  fitting 10000 times, shifting each time the $\Gamma_\mathrm{b}$
  values according to their uncertainties. We estimated the
  probability, $P$, that $\Gamma_\mathrm{b}$ does not decrease with
  radius, as the fraction of iterations giving best-fitting lines with
  non-negative slope values. The $P$ turns out to be less than $1 \%$,
  also in the case where $\rm \log R/R_{e}$, rather than $\rm R/R_{e}$, is
  adopted as independent variable in the fitting. Also, as a further test, 
  we computed a further binned spectrum for NGC\,5557, at a radial position of 
  $\rm R/R_{e}$=0.7, i.e. the outest galactocentric distance explored for NGC\,4552. 
  For this bin, our fitting procedure gives $\Gamma_\mathrm{b}=1.78 \pm 0.2$ (see red triangle in Fig.A2). 
  Including only galactocentric distance less than 0.7 $\rm R/R_{e}$ for NGC\,5557, we still find a 
  probability less than 1 that $\Gamma_\mathrm{b}$ profile does not decraese with radius in this system. 
  Notice that, as discussed in Sec.~\ref{sec:test} for the reference massive galaxy
  NGC\,4552, other ingredients in the stellar population
  modeling (e.g. individual elemental abundances) might contribute to
  further flatten the radial $\Gamma_\mathrm{b}$ trend for
  NGC\,5557. However, the existence of a genuine radial IMF gradient
  for this galaxy is also supported by the Na{\sc I}\,8190 feature, as
  shown in Sec.~\ref{sec:na}, as this feature breaks
  the degeneracy between IMF and other possible effects. Although our
  data suggest that the IMF gradient for NGC\,5557 seems to be shallower 
  than that for NGC\,4552, together with analysis of NGC\,4552 and NGC\,4387, 
  the results for NGC\,5557 further support our conclusions that the enhanced fraction of low-mass stars in massive
  ellipticals is mostly located in their core regions. in their core regions.}

\newpage

\begin{center}
\section{Radial behaviour of additional gravity-sensitive features} \label{app:A}
\end{center}

\subsection{Radial trends of Na{\sc I}\,8190 and NaD} \label{sec:na}
{  The near-infrared Na\ I doublet, at $\lambda\sim$8200\,\AA , is a
  prominent feature in the atmospheres of low-mass dwarves, and it has
  been used, along with the
{Wing-Ford FeH band} 
($\lambda\sim$9900\,\AA), to  derive a significant  excess of low-mass
stars in  the central regions of massive ETGs \citep{vandokkum}.  
Unfortunately, for all the three  galaxies observed with GTC-OSIRIS (Sec.~\ref{sec:data}), the  Na{\sc I}\,8190 
 index  is significantly 
contaminated  by   telluric  absorption.   No   telluric  standard  is
available for the same run as our observations, hampering the study of
{this} 
line.  However,  since the  {\sl relative} absorption  of flux  is virtually the
same  at all  radial positions  for each  galaxy,  
telluric absorption should not affect the amount  of radial variation of the Na{\sc I}\,8190
EW, with only a constant  shift in its absolute value.  Since this
statement  is  only exactly  true  in  case  of no  radial  rotational
velocity,   we have tested it directly  by  constructing two different telluric
absorption models. The ``synthetic'' model (hereafter $TELL1$) is obtained, for each galaxy, 
as the ratio between a  two-SSP model and the
 observed  spectrum  in  the  innermost  radial  bin. The model is obtained by fitting the spectrum, 
 around  the  Na feature, with a linear combination of two SSPs, with age and metallicity being free fitting 
 parameters, and IMF fixed to the result of our fiducial $\chi^2$ minimization procedure in the
 galaxy center (where we obtain consistent results to our previous studies, targeting 
 the central regions of ETGs with a variety of spectral features).
 The second absorption model (hereafter $TELL2$) relies on the spectrum of a 
telluric standard star, observed in a different observing run, but with the same instrumental setup 
as for NGC\,4387 and NGC\,4552. 
For each telluric model, we corrected the 2D galaxy spectra at each radial  position, and re-extracted 
1D spectra for all galacto-centric distance bins, re-measuring 
{the Na{\sc I}\,8190} EWs.
The same kind of analysis was performed for the optical Na doublet,
$NaD$, at $\lambda \sim 5900$~\AA , as well as other IMF-sensitive features (see below). The NaD feature 
turns out to be affected by telluric absorption, and to major extent, by sky emission.
To test also the latter effect, we re-measured both Na features 
by performing sky subtraction with two different methods, i.e. either 
 interpolating the sky from either sides of the slit, far from the galaxy center (hereafter method $SKY1$),
 or (method $SKY2$) performing a second iteration where the 2D spectrum of the galaxy (obtained from $SKY1$) is 
 subtracted off from the original 2D frame, and the sky is re-estimated in a region closer to the galaxy 
 center.

\begin{figure}
\begin{center}
\includegraphics[width=8.5cm]{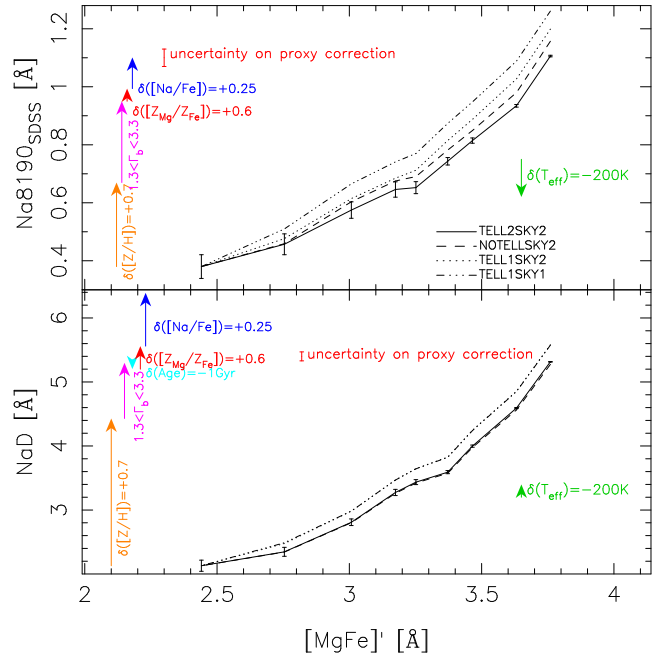}
\end{center}
\caption{ {The Na{\sc I}\,8190 (top panel) and NaD (bottom
    panel) index radial profiles of NGC~4552 are plotted as a function
    of the total metallicity indicator [MgFe]$^\prime$. Black curves
    with different line types (see the black legend in the upper
    panel) correspond to different methods to treat telluric
    absorption, i.e. methods $TELL1$ and $TELL2$ (see the text) and no
    telluric correction ($NOTELL$), as well as different sky
    subtraction procedures (methods $SKY1$ and $SKY2$, see the text).
    Black error bars are the maximum uncertainties (among different
    methods), quoted at the 1\,$\sigma$ level.  In both panels, arrows
    show the expected variation of the indices (for SSP models)
    because of the metallicity (orange), IMF (magenta), age (cyan),
    and [$Z_{Mg}/Z_{Fe}$] (red) gradients for NGC\,4552
    (Sec.~\ref{sec:grid}).  Notice that the effect of varying age is
    not shown for Na{\sc I}\,8190, as it is completely negligible for
    this index.  The effect of varying [Na/Fe] abundance and $\rm
    T_{eff}$ are also shown as blue and green arrows, respectively.
}}
\label{fig:nall}
\end{figure}

Fig.~\ref{fig:nall} compares the radial gradients of Na{\sc I}\,8190 (upper panel) and NaD (lower panel), 
for our reference massive galaxy NGC\,4552,  among different methods used to treat telluric absorption and 
sky emission (see black curves with different line types, as labeled in the upper panel). Since the 
absolute value of the indices was found to be strongly dependent on the 
reduction procedure, we arbitrarely shifted all values, for a given method and each index, to match 
model predictions (from our best-fitting solution, see Sec.~\ref{sec:grid}) in the outermost aperture. 
Notice that the amount of gradients in both indices is fairly consistent (considering the largest quoted 
error bars, i.e. for the 
outermost radial bin), among different methods (even when 
no telluric correction is applied at all), implying that the gradient of both Na features can be safely estimated. 
In both panels, different arrows plot the amount of expected variation  of each index 
-- when moving from the outermost to the innermost 
aperture -- because of the gradients in different stellar 
population properties (metallicity, age, [$\alpha$/Fe], and IMF). 
Summing up all different contributions (orange plus magenta plus red arrows), we predict a radial 
 gradient of $\sim 0.6$~\AA\ and $\sim 3.4$~\AA\ for Na{\sc I}\,8190 and NaD, respectively. From the black curves,
 we see that the observed gradients are $\sim 0.7$--$0.85$~\AA\ for Na{\sc I}\,8190 ($\pm 0.1$~\AA, 
 considering the 
 error bar in the outermost bin, and that on proxy correction) and $\sim 3.3$--$3.5$~\AA\ ($\pm 0.12$) for NaD.
Hence, within the uncertainties, an IMF radial variation reconciles both Na features (once the 
effect of other parameters, in particular that of metallicity, is taken into account). 
The effect of varying the temperature of 
giant stars, $\rm T_{eff}$, in the stellar population models is also illustrated, with green arrows,
in Fig.~\ref{fig:nall}.
Decreasing $\rm T_{\rm eff}$ tends to decrease the model Na{\sc 
I}\,8190, i.e. it goes into an opposite direction to that of a more bottom-heavy IMF. On the other hand,
NaD is insensitive to temperature variations, while TiO bands (panels b-c of Fig.~\ref{fig:grdi}) increase with 
decreasing  $\rm T_{eff}$. Therefore, while the effect of a radially varying IMF is consistent among 
all features (including the Na ones), a radial variation in the temperature is clearly ruled out when
combining optical and NIR features. A similar argument applies to [Na/Fe] abundance (see blue arrows in 
the Figure, showing the effect of increasing [Na/Fe] by $0.25$~dex, as estimated from CvD12 
stellar population models). To mimic the effect of varying $\Gamma_\mathrm{b} $ from $1.3$ to $3.3$ 
(consistent with 
our estimated gradient for NGC\,4552, see Fig.~\ref{fig:tests}), a change of $\sim$0.25~dex in [Na/Fe] 
would be required for NaD.
However, this is far too small (by a factor of two) than that of a varying IMF on Na{\sc I}\,8190
(see magenta and blue arrows in the upper panel of Fig.~\ref{fig:nall}). Hence, an [Na/Fe] gradient 
is not able to explain the observed gradients of both Na features (see, e.g., \citealt{spiniello14}), 
Finally, it is worth noting that alternative explanations for the ``unusual'' strength 
of Na indices  have also been proposed in the literature \citep{Jeong}.

As for NGC\,4552, also for our reference low-mass galaxy, NGC\,4387, we have found that 
the observed Na{\sc I}\,8190 gradients change very mildly, by $\widetilde{<} 0.1$~\AA ,
among different reduction procedures, meaning that the Na{\sc I}\,8190 gradients can be robustly estimated
from our data. Fig.~\ref{fig:na} compares the radial gradients of Na{\sc I}\,8190 (whose EWs are estimated
through methods $TELL2$ and $SKY2$, see above) for the low- and 
high-mass galaxy. A constant  shift is arbitrarily applied  for each galaxy, to  match the
  IMF slope  derived from  our fiducial set of spectral indices (see Sec.~\ref{sec:indices}) 
  in the  outermost radial bins.  
   The Figure also plots the Na{\sc I}\,8190 gradient 
  for the additional massive system, NGC\,5557, observed as part of the same observational campaign as 
  NGC\,4387 and NGC\,4552, but analyzed apart in Appendix~\ref{sec:NGC5557}.
  Remarkably, for the  low-mass ETG,
  no  radial gradient  in Na{\sc  I}8190 is  detected, while  a strong
  decrease is observed  with galactocentric 
  distance for the high-mass galaxies.    
  As shown in Fig.~\ref{fig:na}, the metallicity gradient of NGC\,4552 cannot explain, by itself, 
  the Na{\sc I}\,8190 gradient (as also seen by comparing the size of the orange arrow, to the range 
  of values for the black curves, in the upper panel of Fig.~\ref{fig:nall}). 
  Moreover, the Na{\sc I}\,8190 index is expected
  to decrease with [$\alpha$/Fe] (according to CvD12 stellar population models),  whereas NGC\,4552, 
  as well as NGC\,5557,
  {\sl  both} have high  [$\alpha$/Fe] in  the centre, and  a very  high
  Na{\sc I}\,8190 line-strength.
  In conclusion, the radial gradients of Na  features,   for high-mass  galaxies,  also  point  to  a
  significant radial variation of the stellar IMF, with an excess of
  low-mass   stars   confined to their   central   regions.
 }
  
\begin{figure}
\begin{center}
\includegraphics[width=8.5cm]{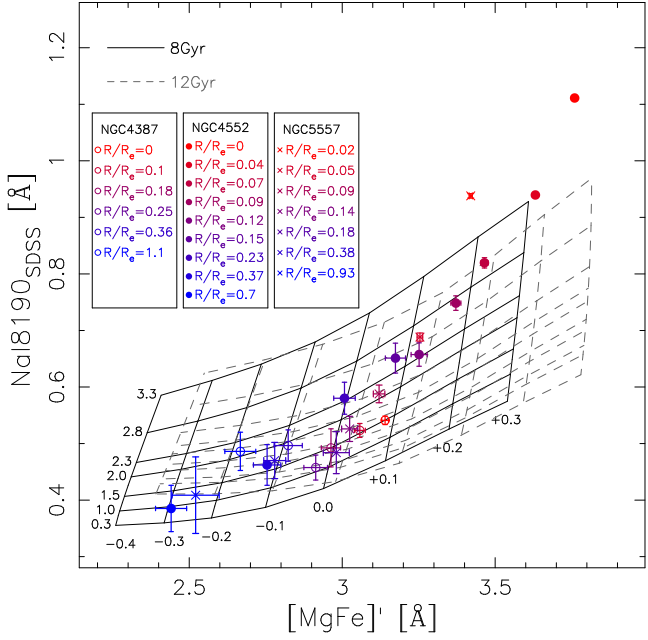}
\end{center}
\caption{  The Na{\sc I}\,8190$_{\rm SDSS}$ line-strength is plotted
  as a function of the total metallicity indicator [MgFe]$^\prime$,
  for NGC\,4387 (empty circles), NGC\,4552 (filled circles), and
  NGC\,5557 (crosses; see Appendix~\ref{sec:NGC5557}). Error bars are
  1\,$\sigma$ statistical uncertainties.  Black and grey grids
  correspond to MILES SSP models with an age of $8$ and $12$\,Gyr
  (i.e. the same as in Fig.~\ref{fig:grdi}), respectively. Notice that
  the Na{\sc I}\,8190 EWs are estimated with method $TELL2SKY2$,
  i.e. the same as for the black solid curve in Fig.~\ref{fig:nall}.
  The strong gradients of Na{\sc I}\,8190$_{\rm SDSS}$ for the
  high-mass galaxies contrast with the flat behaviour of this feature
  for the low-mass system, NGC\,4387.  A constant offset has been
  applied to the line-strengths of each galaxy, to account for the
  uncertainty on the absolute value of the indices, due to telluric
  absorption.}
\label{fig:na}
\end{figure}

\subsection{Radial trends of CaH1 and CaH2} \label{sec:CaH}

\begin{figure}
\begin{center}
\includegraphics[width=8.5cm]{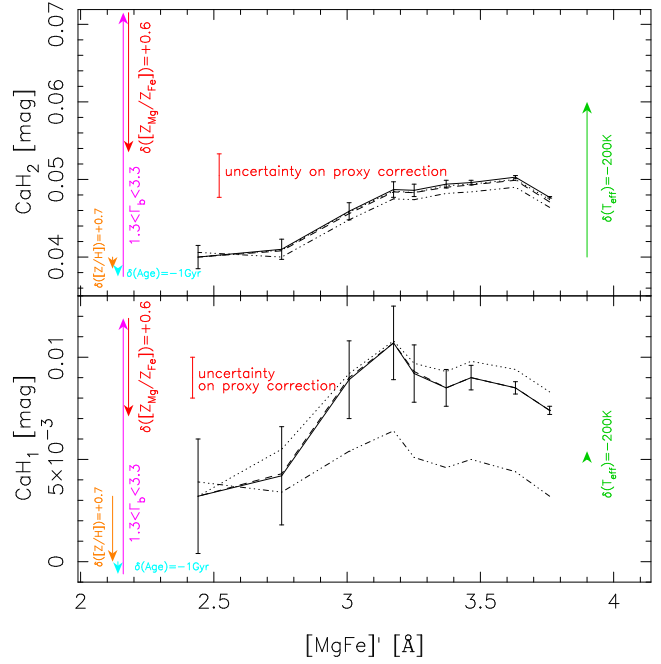}
\end{center}
\caption{{CaH1 (bottom) and CaH2 (top) radial gradients for
    NGC\,4552, i.e. the same as Fig.~\ref{fig:nall} but for CaH,
    rather than Na, features.  The telluric contamination of these
    indices does not allow us to trust the index absolute
    values. However, the radial gradient for CaH2 is robust, while for
    CaH1 we can exclude an increase of the index with radius
    (leftwards in the plot), as it would be the case for a constant
    IMF (see the text).  Notice that CaH2 is fully consistent with our
    results of a varying IMF for NGC\,4552.  In general, because of
    the different sensitivity of the two indices to a variation in
    $\rm T_{eff}$ (green arrows), one cannot explain the behaviour of
    both CaH features without a varying IMF with radius.}}
\label{fig:CaH}
\end{figure}

{ We discuss here the radial behaviour of the two IMF-sensitive
  indices, CaH1 and CaH2, recently proposed by~\citet{Spiniello2013}.
  Both features (in particular CaH2) are affected by telluric
  absorption in the spectra of our reference high mass galaxy,
  NGC\,4552. However, as done for Na features, we compare the radial
  gradients of both features among different reduction procedures, in
  order to gain further insights into the radial variation of stellar
  population properties for our reference massive galaxy.
  Fig.~\ref{fig:CaH} is similar to Fig.~\ref{fig:nall}, but plotting
  CaH2 (top) and CaH1 (bottom), rather than Na features, as a function
  of [MgFe]$^\prime$ (both CaH features have been arbitrarely shifted
  in the Figure).  As seen by the size of the cyan arrows, both
  indices are insensitive to age, consistent with Spiniello et
  al.(2014a, 2014b).  CaH2 is also insensitive to total metallicity,
  while CaH1 shows a mild trend to decrease with increasing
  [Z/H]. Notice that according to our empirical approach (LB13), both
  indices (in particular CaH2) tend to {\it decrease} significantly
  with $\rm [Z_{Mg}/Z_{Fe}]$ (i.e. $[\alpha/Fe]$; see red arrows), at
  fixed total metallicity.  As seen by summing up the orange, magenta,
  and red arrows in the upper panel of Fig.~\ref{fig:CaH}, the radial
  behaviour of CaH2 is fully consistent with our results, i.e. a radial
  decrease of IMF slope in NGC\,4552. For CaH1, the situation is more
  uncertain, as this index turns out to be also significantly affected
  by airglow (see dot-dashed curve), and thus its radial gradient is
  either null or negative (i.e. decreasing outwards in the galaxy,
  leftwards in Fig.~\ref{fig:CaH}), depending on the reduction
  procedure. We notice that, for a fixed IMF slope, the CaH1 gradient
  in NGC\,4552 should be positive, as both metallicity and $\rm
  [Z_{Mg}/Z_{Fe}]$ increase towards the galaxy center, CaH1 decreasing
  with both paramaters. Hence, regardless of the reduction procedure,
  our results for CaH1 also imply a radial decrease of IMF slope with
  radius in our reference massive galaxy. We notice that the radial
  gradient of CaH2 might also be explained by decreasing $T_{eff}$ in
  the models by $\sim 300$~K. However, this would be inconsistent, at
  fixed IMF, with the radial trend of CaH1, as the latter is
  insensitive to $T_{eff}$. In conclusion, also CaH features imply an
  IMF gradient for NGC\,4552, ruling out a radial variation of
  $T_{eff}$.  }

\subsection{Radial trend of aTiO} \label{sec:aTiO}

\begin{figure}
\begin{center}
\includegraphics[width=8.5cm]{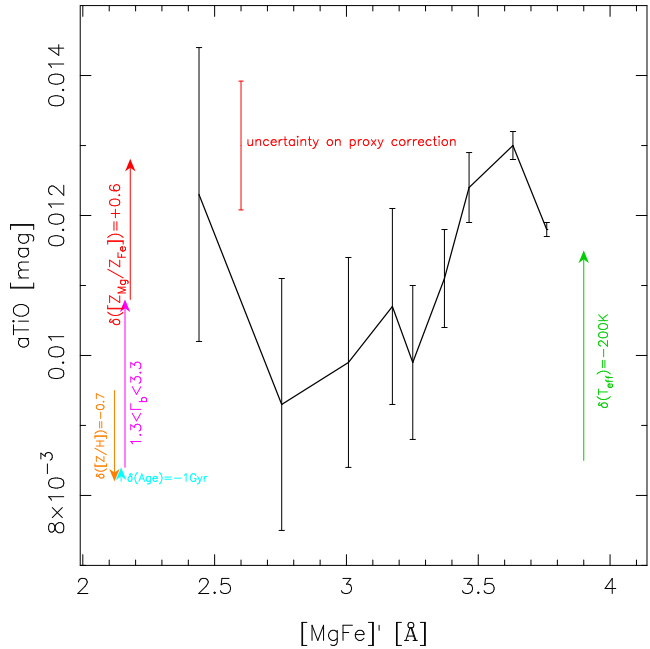}
\end{center}
\caption{{Radial profile of the interpolated aTiO index for NGC\,4552 (see the text). 
Notice the large error bars, in the outer radial bins (left in the plot) preventing us to use 
this index to significantly constrain the IMF gradient. 
Notice that because the index is not affected by telluric absorption,
while airglow contamination is avoided by our interpolation procedure, no difference exists
among different reduction procedures and we did not apply any rigid shift to the line 
strengths (as, e.g., in Figs.~\ref{fig:nall} and~\ref{fig:CaH}).
}}
\label{fig:aTiO}
\end{figure}

{ Finally, we discuss the radial behaviour of the aTiO
  IMF-sensitive feature ($\lambda \sim 5500$~\AA ), also proposed by
  \citet{Spiniello2013}. Being present in the atmospheres of both
  dwarf and giant stars, the aTiO is a potentially good IMF
  indicator. In our data, a strong sky emission line ($\lambda \sim
  5780$~\AA ) lays within the central bandpass of the index. To
  overcome this problem, we have linearly interpolated both data and
  models over the affected region.  The radial profile of aTiO for
  NGC\,4552 is shown in Fig.~\ref{fig:aTiO}. In this case, since the
  feature is not affected by telluric absorption, we do not apply any
  shift to the line strenths. Also, since there is no difference among
  different reductions, only one black curve is shown in the Figure.
  Unfortunately, because of the large error bars, the aTiO has little constraining
  power on the radial variation of the IMF for NGC\,4552, and thus we have
  not included it in our $\chi^2$ minimization procedure. 
  }

\newpage

\begin{center}
\section{Correlated uncertainties on IMF and other parameters} \label{sec:corre}
\end{center}

\begin{figure}
\begin{center}
\includegraphics[width=8.5cm]{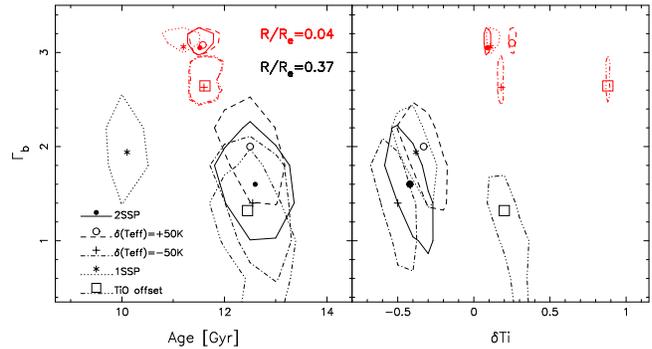}
\end{center}
\caption{Probability density contours of  IMF slope vs. Age (left) and
  IMF slope vs.  $\delta$Ti residual correction (right) for the second
  innermost (red) and outermost  (black) radial bins of NGC\,4552. The
  contours  correspond  to  1$\sigma$  confidence  levels.   Different
  line-styles  are  the  results   of  different  tests,  varying  the
  modelling/fitting approach.   The contours in the left  panel show a
  mild correlated  variation, illustrating  how our method  breaks the
  $Age$--IMF  slope degeneracy.   The right  panel shows  a correlated
  variation  of IMF  slope  and $\delta$Ti.   However,  this trend  is
  almost  orthogonal to  the radial  IMF-slope gradient,  proving that
  [Ti/Fe] cannot be responsible for the observed TiO gradients.}
\label{fig:cont}
\end{figure}

In Fig.~\ref{fig:cont}, we show the PDF contours of our best-fitting
solution for the second innermost and outermost radial bins of
NGC\,4552, in the $\Gamma_b$--$Age$ (left) and $\Gamma_b$--$\delta$Ti
(right) diagrams. Different line types show the impact of a number of
effects, i.e.  changing the modelling and fitting approach, as well as
flux calibration issues (see details in \S\ref{sec:test}), on our
results.  Notice that in all cases, a clear IMF radial gradient for
NGC\,4552 is detected. The left panel shows that there is very little
correlation of best-fitting $Age$ and $\Gamma_b$ values, i.e.  our
approach is able to break the $Age$ vs.  IMF slope degeneracy
affecting spectral fitting at optical wavelengths alone.  As seen from
the black contours in the $\Gamma_b$--$\delta$Ti diagram
(corresponding to the outermost radial bin of NGC\,4552), the errors
on $\Gamma_b$ and $\delta$Ti are significantly correlated, but the
direction of this correlation is almost orthogonal to the radial
variation of best-fitting $\Gamma_b$ and $\delta$Ti values.  This
ultimately results from the fact that the response of both TiO$_1$ and
TiO$_2$ molecular features to [Ti/Fe] is
comparable~\citep{johansson12} (i.e.  $C_{TiO1} \sim C_{TiO2}$ in
Eq.~1), while the observed radial gradient of TiO$_2$ in NGC\,4552 is
about twice that of TiO$_1$. 
{It is worth noticing the strong dependence of $\delta$Ti on
  flux calibration, as shown by the red and black contours associated
  to the squares in the right panel. These contours are obtained after
  applying an artificial shift to TiO1 and TiO2, mimicking a possible
  uncertainty on flux calibration (see \S\ref{sec:test}). This test
  shows that the absolute value of $\delta$Ti is not trustable, but it
  might be significantly affected by the reduction procedure.
On the other hand, the $\delta$Ti gradient
is much more robust ($\Delta \delta$Ti $\sim 0.4$~dex, 
for all different tests),  and might be tracing 
a true radial variation of [Ti/Fe].}

\label{lastpage}

\end{document}